\documentclass[]{pasj01}
\usepackage{subfigure}
\usepackage{natbib}

\newcommand{\nngc}{$\nu^2$GC}

\newcommand{\HI}{H\emissiontype{I} }

\newcommand{\Msun}{M_{\odot}}

\begin{document} 
\Received{}
\Accepted{}

\title{The New Numerical Galaxy Catalog ($\nu^2$GC): An Updated Semi-analytic Model of Galaxy and AGN with Large Cosmological N-body Simulations}

\author{Ryu \textsc{Makiya}\altaffilmark{1}}
\altaffiltext{1}{Institude of Astronomy, University of Tokyo, 2--21--1 Osawa,
   Mitaka-shi, Tokyo 181-0015}
\email{makiya@ioa.s.u-tokyo.ac.jp}

\author{Motohiro \textsc{Enoki}\altaffilmark{2}}
\altaffiltext{2}{Faculty of Business Administration, Tokyo Keizai University, 
Kokubunji, Tokyo, 185-8502, Japan}

\author{Tomoaki \textsc{Ishiyama}\altaffilmark{3}}
\altaffiltext{3}{Institute of Management and Information Technologies, Chiba University, 
1-33, Yayoi-cho, Inage-ku, Chiba, 263-8522, Japan}

\author{Masakazu A.R. \textsc{Kobayashi}\altaffilmark{4}}
\altaffiltext{4}{
Research Center for Space and Cosmic Evolution, Ehime University, Matsuyama, Ehime, 790-8577, Japan}

\author{Masahiro \textsc{Nagashima}\altaffilmark{5,6}}
\altaffiltext{5}{
Faculty of Education, Bunkyo University, Koshigaya, Saitama 343-8511, Japan}
\altaffiltext{6}{
Faculty of Education, Nagasaki University, 1-14, Bunkyo-machi, Nagasaki City, Nagasaki, 852-8521, Japan
}

\author{Takashi \textsc{Okamoto}\altaffilmark{7}}
\altaffiltext{7}{
Department of Cosmosciences, Graduates School of Science, Hokkaido University, N10 W8, Kitaku, Sapporo 060-0810, Japan}

\author{Katsuya \textsc{Okoshi}\altaffilmark{8}}
\altaffiltext{8}{
Tokyo University of Science, Oshamambe, Hokkaido, 049-3514 Tokyo, Japan}

\author{Taira \textsc{Oogi}\altaffilmark{5,6}}

\author{Hikari \textsc{Shirakata}\altaffilmark{7}}

\KeyWords{cosmology: theory --- galaxies: evolution --- galaxies: formation 
--- methods: numerical}
\maketitle

\begin{abstract}
We present a new cosmological galaxy formation model, $\nu^2$GC, as an
updated version of our previous model $\nu$GC. 
We adopt the so-called ``semi-analytic'' approach, 
in which the formation history of dark matter halos is computed by {\it N}-body 
simulations, while the baryon physics such as gas cooling, star formation and 
supernova feedback are simply modeled by phenomenological equations. 
Major updates of the model are as follows: 
(1) the merger trees of dark matter halos are constructed in state-of-the-art 
{\it N}-body simulations,
(2) we introduce the formation and evolution process of supermassive 
black holes and the suppression of gas cooling due to active galactic
nucleus (AGN) activity, 
(3) we include heating of the intergalactic gas by the cosmic UV 
background, 
and (4) we tune some free parameters related to the astrophysical 
processes using a Markov chain Monte Carlo
method.
Our {\it N}-body simulations of dark matter halos have unprecedented 
box size and mass resolution (the largest simulation contains 550 billion 
particles in a 1.12 Gpc/h box), enabling the study of much smaller and rarer objects.
The model was tuned to fit the luminosity functions of local galaxies 
and mass function of neutral hydrogen. 
Local observations, such as the Tully-Fisher relation, size-magnitude 
relation of spiral galaxies and scaling relation between the bulge mass and 
black hole mass were well reproduced by the model.
Moreover, the model also well reproduced the cosmic star formation 
history and the redshift evolution of rest-frame {\it K}-band luminosity 
functions. The numerical catalog of the simulated galaxies and AGNs is publicly 
available on the web. 
\end{abstract}

\section{Introduction}
Understanding the formation and evolution of galaxies is
a primary goal in astrophysics.
Over the past decades, wide and deep surveys at various wavelengths
have acquired numerous observational data of galaxies 
spanning a wide range of galaxy types, magnitudes and distances
\cite[see][for review]{Madau2014}.
Theoretically, the $\Lambda$ cold dark matter (CDM) paradigm 
can explain the formation of the large scale structures governed by 
dark matter (DM) and dark energy.
However, at the scale of galaxies, where baryons play important roles,
several inconsistencies remain between the theory and observations.
To fully elucidate galaxy formation, we need to solve the 
complicated physical processes of baryons within the framework of $\Lambda$-CDM universe.

One of the most promising ways to address this issue is 
the hydrodynamical simulations of cosmological galaxy formation,
in which the equations of gravity, hydrodynamics, and thermodynamics
are solved self-consistently.
However, the mass resolution and box size of these simulations are
still limited by computational costs, 
and the physical processes on scales smaller than 
the numerical resolution are treated by phenomenological 
recipes (the so-called ``sub-grid physics''),
which contain large uncertainties \citep[see][for review]{Springel2012}.

``Semi-analytic models'' (SA models) are also widely used 
in studies of cosmological galaxy formation
(e.g., \citealt{Kauffmann1993b, Cole1994, Cole2000, Somerville1999}).
In SA models, the formation and evolution history of 
dark matter halos are explicitly modeled by analytical formulae
or {\it N}-body simulations, while the complicated baryon physics 
are modeled by phenomenological equations.
The advantage of this technique is its lower computational cost
than numerical simulation, enabling us to create a large
sample of mock galaxies covering the wide range of physical properties 
such as mass, magnitude, and spatial scale.
We can also investigate a wide range of the parameter space and 
test various models of the baryon physics.
However, to discuss the galaxy-scale dynamics, we need to combine SA models 
(which do not explicitly treat such dynamics) with fully numerical simulations. 
See, e.g., \cite{Somerville2014} for more detailed review of the physical 
models of cosmological galaxy formation.

In this paper we introduce our new galaxy formation model, 
{\it New Numerical Galaxy Catalog} ($\nu^2$GC), an updated version
of {\it Numerical Galaxy Catalog} ($\nu$GC) presented in 
\citet[hereafter N05; see also \citealt{Nagashima2004}]{Nagashima2005}.
Our model is an SA model,
in which we directly extract the merger trees of DM halos from {\it N}-body simulations, following the pioneering work of \cite{Roukema1997}.
The $\nu$GC model and its variants have been used in many studies
(e.g., \citealt{Kobayashi2007,Kobayashi2010,Okoshi2010,Makiya2011,Makiya2014,
Enoki2014,
Shirakata2015,Oogi2016}).
Major updates of the new model from the version of N05 are as follows: 
(1) $\nu^2$GC adopts the new {\it N}-body simulations of DM halos recently 
presented by \cite{Ishiyama2015}, (2) the formation and evolution process of 
supermassive black holes (SMBHs) and suppression of gas cooling by active 
galactic nuclei (AGNs) are included, (3) heating of the intergalactic gas 
by the cosmic UV background is included,
and (4) some parameters are tuned to fit the local luminosity
functions and \HI mass function using a Markov chain Monte Carlo 
(MCMC) method.

Several other groups have also proposed SA models \citep[see][for review]{Somerville2014}.
Each of these models is based on different {\it N}-body simulations 
and adopts different equations of the baryon physics.
For a comparison study of different galaxy formation models, 
see \cite{Knebe2015}.
Our model is characterized by the substantially higher mass resolution 
of the {\it N}-body simulations of DM halos, comparing with 
other large box simulations.
Our simulations consist of seven runs with varying mass resolutions and
box sizes, as listed in table~\ref{tb:nbody1}.
For example, the largest simulation, $\nu^2$GC--L, includes $8192^3$ DM 
particles in a box of 1.12 $h^{-1}$~Gpc, and the minimum halo mass reaches
$8.79 \times 10^9 M_{\odot}$. Comparing with the Millennium simulation 
(\citealt{Springel2005}), 
the $\nu^2$GC--L simulation is four times better in mass resolution and is 
11 times larger in spacial volume.
The $\nu^2$GC-H2 simulation has the highest mass resolution among our simulations.
The minimum halo mass reaches $1.37\times10^8~\Msun$, below the effective Jeans mass 
at high redshift (N05). 
This mass resolution is 2 times better than Millennium-II
simulation (\citealt{Boylan-Kolchin2009}), although the spatial volume of $\nu^2$GC--H2
is 3 times smaller than that of Millennium-II.
These high mass resolution and large spatial volume enable us to obtain
a statistically significant number of mock galaxies and AGNs, 
even at high redshifts.
Moreover, we adopt the cosmological parameters
recently obtained by the Planck satellite \citep{Planck2014}, 
while the most of other SA models are based on the parameters obtained by
Wilkinson microwave anisotropy probe (WMAP), which significantly
differ from the Planck results.
For more detailed comparison with other cosmological {\it N}-body
simulations, see \cite{Ishiyama2015}.

This paper describes the basic properties of our model,
focusing on the nature of local galaxies.
The properties of distant galaxies and AGNs will be discussed in 
our forthcoming papers.

The paper is organized as follows.
Sections \ref{sec:model} and \ref{sec:DetParams} present the details of our
model and the parameter fitting method, respectively.
The general properties of our numerical galaxy catalog are presented in 
section \ref{sec:catalog}, and sections \ref{sec:localresults} and 
\ref{sec:highzresults} compare the model predictions with the observed 
properties of local and distant galaxies. 
Section \ref{sec:summary} summarizes the paper.
The mock galaxy catalog produced by the our new model 
is publicly available on the web\footnote[1]{http://www.imit.chiba-u.jp/faculty/nngc/}.

\section{Model descriptions}
\label{sec:model}

\begin{figure*}
    \begin{center}
    \includegraphics[width=2\columnwidth]{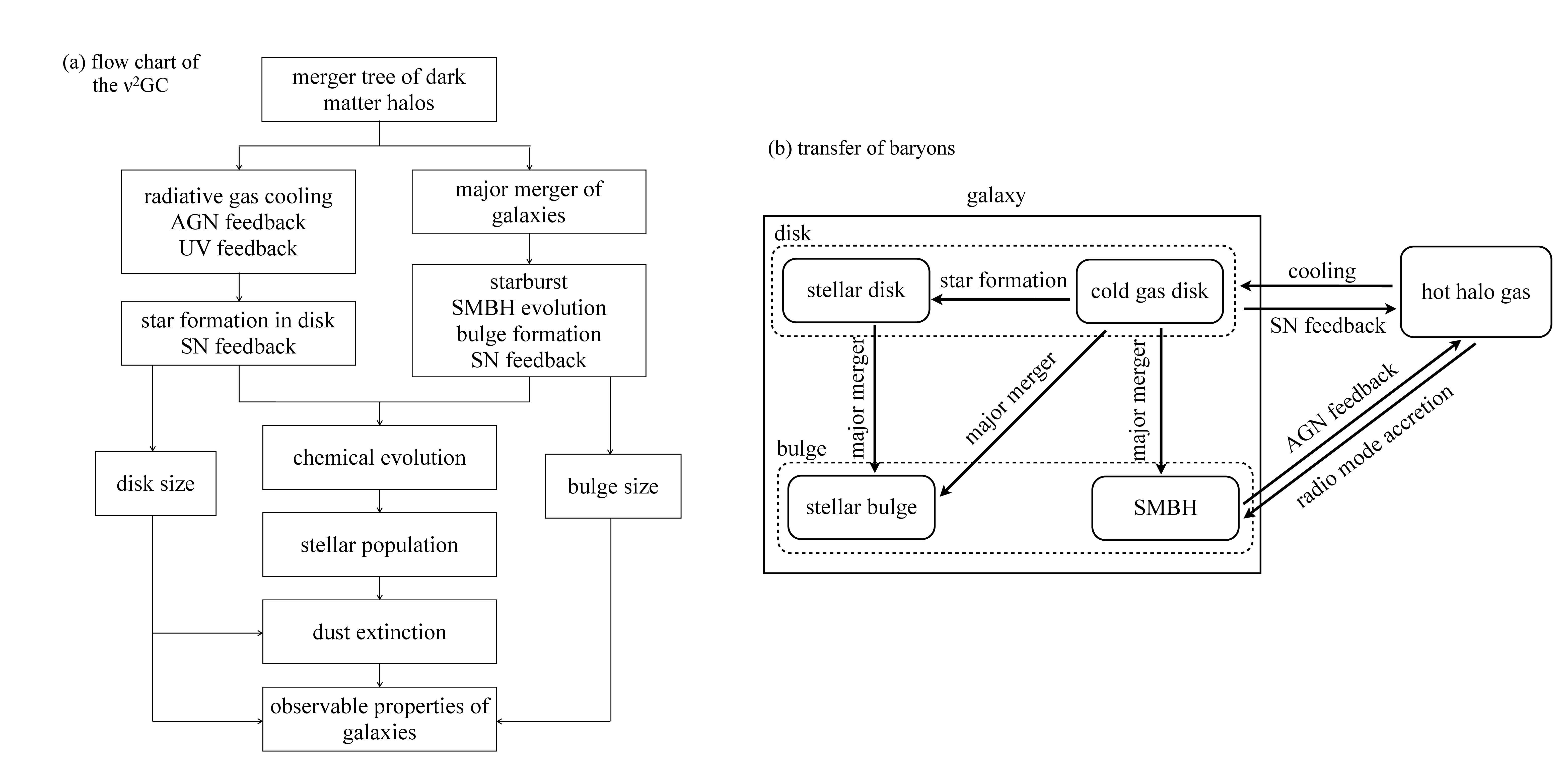}
    \end{center}
    \caption{
    Schematics of the model. (Left) Flow chart of the model showing how the model predicts
    the observable properties of galaxies. (Right) Schematic of the transfer of baryon components.
    }
    \label{fig:chart}
\end{figure*}

In the CDM universe, DM halos hierarchically grow from small to 
large scales.
When a DM halo collapses, the contained gas is heated to virial temperature 
by shock, and then gradually cools by radiative cooling
(in reality, a gas in low-mass halos would not be shock heated but directly 
forms cold gas disk; see section~\ref{sec:Gascooling} for more detailed 
discussion).
The cooled gas condense into stars; those stars and dense cold gas constitute 
galaxies.
The massive stars formed by this process explode as supernovae (SNe),
blowing out surrounding cold gas. 
This process suppresses further star formation (the so-called ``SN feedback'').
Massive stars also eject metals.
Galaxies in a common DM halo sometimes merge into more massive galaxies, 
and galaxy bulge is formed as a merger remnant;
cold gas in the merger remnant is converted into stars with short timescale, a phenomenon called a starburst. 
During the starburst, a fraction of the cold gas gets accreted by the supermassive 
black hole (SMBH) at the galaxy center.
By repeating these processes, galaxies and SMBHs have formed 
and evolved to the present epoch.
Each of these processes is described in the following subsections.
Figure \ref{fig:chart} displays an overview of the model.

\subsection{Dark matter merger trees}
The merger trees of DM halos are directly extracted from a series of large
cosmological {\it N}-body simulations, called the \nngc\ simulations
\citep{Ishiyama2015}.  The basic properties of the \nngc\ simulations are
summarized in table \ref{tb:nbody1}.  We conducted seven
simulations, varying the mass resolution and spatial volume.  
The largest \nngc-L run simulated the motions of $8192^3$ (550 billion) DM
particles in a comoving box of $1.12 \, h^{-1} \rm Gpc$.
The mass resolution was $2.20 \times 10^{8} \, h^{-1}
M_{\odot}$, which is the best one among
simulations applying boxes larger than $1 \, h^{-1} \rm Gpc$.  The
mass resolution of the run with the smallest box (\nngc-H2) was $3.44
\times 10^{6} \, h^{-1} M_{\odot}$, which is sufficient to resolve
small dwarf galaxies.  By combining these simulations, we can 
generate mock catalogs of galaxies and AGNs with unprecedentedly
high resolution and statistical power.

The cosmological parameters of the \nngc\ simulations were based on the
concordance $\Lambda$CDM model consistent with observational results
obtained by the Planck satellite \citep{Planck2014}. Namely,
$\Omega_0=0.31$, $\Omega_b=0.048$, $\lambda_0=0.69$, $h=0.68$,
$n_s=0.96$, and $\sigma_8=0.83$.  The \nngc\ simulations were
conducted by using a massively parallel TreePM code GreeM
\citep{Ishiyama2009, Ishiyama2012}.  DM halos are identified by the
friends-of-friends (FoF) group finder \citep{Davis1985}, with the linking
parameter $b=0.2$.  The smallest halos consisted of 40 particles.  The
spatial positions of the halos were tracked by using those of the most
bound particles.  
It has been known that the properties of halo merger tree are depend on the halo finding algorithm and tree building algorithm 
(see, e.g., \citealt{Knebe2011,Onions2012,Elahi2013,Srisawat2013,Avila2014,Lee2014}).
For further details of the \nngc\ simulations and
the method for extracting the merger trees, see companion paper,
\citet{Ishiyama2015}.

\begin{table*}[t]
\caption
{Details of the \nngc\ simulations. {\it N} is the number of simulated
  particles, $L$ is the comoving box size, $m$ is the particle mass
  resolution, $M_{\rm min}$ is the mass of the smallest halos, the
  total number of halos, and $M_{\rm max}$ is the mass of the largest
  halo in each simulations.  The smallest halos consist of 40
  particles.  In the last two columns, values at z=0 are presented
  except for the \nngc-H3 simulation, which was
  stopped at $z=4$. }\label{tb:nbody1}
  \centering
\begin{tabular}{lcccccc}
\hline
Name & $N$ & $L(h^{-1} \rm Mpc)$ & $m (h^{-1} M_{\odot})$ & $M_{\rm min} (h^{-1} M_{\odot})$ & \#Halos & $M_{\rm max} (h^{-1} M_{\odot})$\\
\hline
\nngc-L  & $8192^3$ & $1120.0$ & $2.20 \times 10^{8}$ & $8.79 \times 10^{9}$ & $421,801,565$ & $4.11 \times 10^{15}$\\
\nngc-M  & $4096^3$ & $560.0$  & $2.20 \times 10^{8}$ & $8.79 \times 10^{9}$ & $52,701,925$ & $2.67 \times 10^{15}$\\
\nngc-S  & $2048^3$ & $280.0$  & $2.20 \times 10^{8}$ & $8.79 \times 10^{9}$ & $6,575,486$ & $1.56 \times 10^{15}$ \\
\nngc-SS  & $512^3$ & $70.0$  & $2.20 \times 10^{8}$ & $8.79 \times 10^{9}$ & $103,630$ & $6.58 \times 10^{14}$ \\
\nngc-H1 & $2048^3$ & $140.0$  & $2.75 \times 10^{7}$ & $1.10 \times 10^{9}$ & $5,467,200$ & $4.81 \times 10^{14}$  \\
\nngc-H2 & $2048^3$ & $70.0$   & $3.44 \times 10^{6}$ & $1.37 \times 10^{8}$ & $4,600,746$ & $4.00 \times 10^{14}$ \\
\hline
\nngc-H3 & $4096^3$ & 140.0  & $3.44 \times 10^{6}$ & $1.37 \times 10^{8}$ & $44,679,543 (z=4)$ & $1.15 \times 10^{13} (z=4)$\\
\hline
\end{tabular}
\end{table*}

\subsection{Gas cooling}
\label{sec:Gascooling}
We define the formation epoch of the DM halo as 
the time at which the DM halo mass doubles its mass 
since the last formation epoch \citep{Lacey1993}.
At this time, the physical quantities of the halo, such as circular velocity,
halo age and mass density are re-estimated.
Before reionization of the universe, the mass fraction of 
the baryonic matter in a collapsing DM halo is given by
$\langle f_\mathrm{b} \rangle \equiv \Omega_{\rm b}/\Omega_{0}$
(after cosmic reionization, the baryon mass in a halo
deviates from $\langle f_\mathrm{b} \rangle$; 
see subsection \ref{subsec:UVback}).
The baryonic matter consists of diffuse hot gas, dense cold gas, stars,
and black holes.
When a mass of DM halo decreases, diffuse hot gas also decreases
at the same ratio with the decrease of DM mass,
while a mass of other baryon components does not change.

When a DM halo of circular velocity $V_{\rm circ}$ forms, the contained gas
is shock heated to the virial temperature $T_{\rm vir}$ of the halo:
\begin{equation}
 T_{\rm vir}=\frac{1}{2}\frac{\mu m_{\rm p}}{k_{\rm B}}V_{\rm circ}^{2},
\end{equation}
where $m_{\rm p}, k_{\rm B}$ and $\mu$ are the proton mass,
Boltzmann constant and mean molecular weight, respectively.
Following \citet{Shimizu2002}, we assume that the hot gas distributes 
through the DM halo with an isothermal density profile 
with a finite core radius:
\begin{equation}
 \rho_{\rm hot}(r)=\frac{\rho_{\rm hot,0}}{1+(r/r_{c})^{2}},
\end{equation} 
where $r_{c}=0.22R_{\rm vir}/c$, and $R_{\rm vir}$ is the virial radius of
the host halo. 
The concentration parameter $c$ is known to be a function of DM halo mass and redshift. 
We used the analytical formula of $c$ proposed by \citet{Prada2012}, which 
is obtained by fitting cosmological $N$-body simulations.
The model of \citet{Prada2012} and \citealt{SanchezConde2014} 
are consistent with our $N$-body simulations.

After the collapse of a DM halo, the hot gas gradually cools via radiative cooling,
forming a cold gas disk at the halo center.
Stars born from the condensed cold gas;
and a stellar disk and a cold gas disk consist a galaxy (see section 
\ref{sec:sffb}).
The rate of gas cooling is calculated following the model
proposed by \cite{White1991}, which is adopted in most SA models.
The time scale of radiative cooling, $t_{\rm cool}$, is calculated as 
\begin{equation}
 t_{\rm cool}(r)=\frac{3}{2}\frac{\rho_{\rm hot}(r)}{\mu m_{\rm p}}
  \frac{k_{\rm B}T_{\rm vir}}{n_{\rm e}^{2}(r)\Lambda(T_{\rm vir}, Z_{\rm hot})},
  \label{eq:tcool}
\end{equation}
where $n_{\rm e}(r)$ is the electron density of hot gas at $r$, $Z_{\rm
hot}$ is the metallicity of hot gas, and $\Lambda$ is a
metallicity-dependent cooling function provided by \citet{Sutherland1993}. 
In each time step, the hot gas within the {\it cooling radius}
cools and accretes onto the central cold gas disk.
The cooling radius $r_{\rm cool}(t)$ is defined as the radius at 
which the cooling time scale is equal to the time elapsed since 
the halo formation epoch.
If the cooling radius exceeds the virial radius $R_{\rm vir}$, 
we set to $r_{\rm cool} = R_{\rm vir}$.
In this case the mass accretion rate of cold gas should be limited by
the free fall time, rather than the cooling time.
However, we set the time step to be 
comparable to the dynamical time scale of halo at each epoch, 
thus this could not cause a serious problem.

We further assumed that the radial profile of hot gas is kept unchanged 
until the DM halo doubles its mass, allowing the existence of ``cooling hole''
at the centre of the halo (i.e., no hot gas is distributed at $r < r_{\rm cool}$).
This assumption is clearly unphysical, thus the effect on the gas cooling rate 
should be checked.
\cite{Monaco2014} compared their SA model, {\sc MORGANA} (\citealt{Monaco2007}),
with other SA models and hydrodynamical simulations to test the cooling models.
In their model, the cooling radius is treated as a dynamical variables and
the gas profile is recomputed in each time step.
They also consider the pressure balance between the hot gas and cooled gas,
which determines the size of the cooling hole.
One of other SA models examined in \cite{Monaco2014} adopts the cooling model of
\cite{White1991}, as well as our model.
\cite{Monaco2014} show that the different cooling models adopted in SA models 
only makes a marginal difference in cooling rate.
See also \cite{DeLucia2010} for a test of the cooling models.

As shown in equation (\ref{eq:tcool}), the cooling time scale depend on 
both the temperature and metallicity of the gas.
In our model, the chemical enrichment of the hot gas due to the star formation and
SN feedback is consistently solved as shown in subsection 
\ref{sec:sffb}.

Note that the above assumption that the hot gas is heated up by shock
at collapse of host halos is adopted just for simplicity.  In reality,
the cooling time scale of hot gas within galactic scale halos is much
shorter than their dynamical time scale.  Therefore, the hot gas should
cool immediately rather than spherically re-distributing throughout the 
host halos.  In any case, because the cooling time scale is very
short, almost all the hot gas cools and thus our assumption is
expected to work well.  For its opposite case, within cluster-scale
halos,  the cooling time scale is very long owing to the high virial
temperature and the AGN feedback.  Again the assumption should be
good.  For the intermediate scale, we might need more sophisticated
treatment.  Along with the AGN feedback, these process should be improved in 
future versions of the model.

\subsection{Photoionization heating due to an UV
  background}\label{subsec:UVback}

Intergalactic gas is photo-heated by the cosmological UV radiation field 
produced by galaxies and quasars.  
Because the heated gas cannot be accreted by small halos with shallow 
gravitational potential wells, 
photo-heating quenches star formation in small galaxies and hence decreases
the number densities of dwarf galaxies (e.g., \citealt{Doroshkevich1967}; 
\citealt{Couchman1986}).  The characteristic halo mass $M_\mathrm{c}$,
below which a halo cannot retain the heated gas, has been 
investigated by using cosmological hydrodynamic simulations (e.g., 
\citealt{Gnedin2000}).

In this context, \cite{Okamoto2008} performed high-resolution
cosmological hydrodynamical simulations with a time-dependent UV
background radiation field.  They found that the redshift evolution of
the characteristic mass, $M_\mathrm{c}(z)$, is determined by the following
factors for each halo: the relation between $T_\mathrm{vir}$ and the
equilibrium temperature for the gas $T_\mathrm{eq}$ at the edge of the halo, 
at which the density can be approximated as one third of the cosmic mean, 
and its merging history (see section~4 in \citealt{Okamoto2008}). 
They also found that the mass fraction of baryonic matter 
in halos with mass $M_{\rm h}$ at redshift $z$ is well-fitted by 
the following formula, originally proposed by \cite{Gnedin2000}:
\begin{eqnarray}
 f_\mathrm{b} (M_{\rm h}, z) &=& 
\langle f_\mathrm{b} \rangle \nonumber \\
&& \times \left\{1 + (2^{\alpha_\mathrm{UV}/3} - 1) \left[\frac{M_{\rm h}}{M_\mathrm{c} (z)}\right]^{-\alpha_\mathrm{UV}}  \right\}
 ^{-\frac{3}{\alpha_\mathrm{UV}}},
  \label{eq:fb-O08}
\end{eqnarray}
where the parameter $\alpha_\mathrm{UV}$ controls the rate of decrease of
$f_\mathrm{b}$ in low-mass halos, here set to
$\alpha_\mathrm{UV} = 2$.  While $f_\mathrm{b} (M_{\rm h}, z)$ equals to
$\langle f_\mathrm{b} \rangle$ for the halos with $M_{\rm h} \gg
M_\mathrm{c}(z)$, it goes to zero in proportion to $(M_{\rm h} / M_\mathrm{c})^3$ for
the halos with $M_{\rm h} \ll M_\mathrm{c}(z)$.
This decrease is attributed to the suppressed accretion of
photo-heated baryonic matter onto the halos.
This prescription, given by \cite{Okamoto2008},
is newly incorporated into our \nngc\ model.  Although all the above factors that
determine $M_\mathrm{c}(z)$ are evaluable in our
\nngc\ model, we simply adopt their resultant $M_\mathrm{c}(z)$ itself
in order to avoid a relatively large computational cost to obtain
$T_\mathrm{eq} (\langle f_\mathrm{b} \rangle \rho_\mathrm{vir} / 3)$.

\begin{figure}
    \includegraphics[width=\columnwidth]{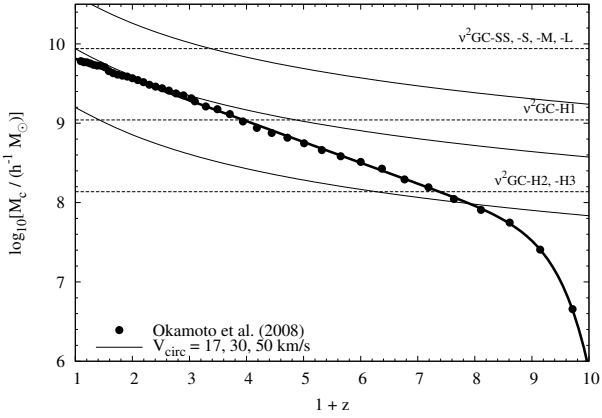}
    \caption{
    The redshift evolution of the characteristic mass $M_{\rm c}$, 
    below which a halo cannot retain the intergalactic gas 
    due to the heating by the     cosmological UV radiation field.
    The black filled circles show the results of the cosmological 
    hydrodynamical simulations performed by \cite{Okamoto2008}.
    The thick solid line presents the fitting formula described in
    equation~(\ref{eq:fiteq}).
    The horizontal dashed lines denote the minimum halo mass of
    our {\it N}-body simulations ($M_{\rm min} = 8.79\times10^{9}, 
    1.10\times10^{9}$ and $3.44\times10^{8}~\Msun$ from top to 
    bottom,
    respectively).
    The thin solid lines correspond to the fixed circular velocity
    $V_{\rm circ} = 50$, $30$ and $17$~km~s$^{-1}$
    from top to bottom.
    }
    \label{fig:Mcz}
\end{figure}

The details how to incorporate the prescription of \cite{Okamoto2008}
 are as follows.  Before reionization, which is assumed to
instantaneously occur at $z = z_\mathrm{reion}$, all halos contain
baryonic matter with a mass fraction of $f_\mathrm{b} = \langle
f_\mathrm{b} \rangle$ regardless of their masses, as described in
subsection~\ref{sec:Gascooling}.  
After reionization, the expected
baryon fraction $f_\mathrm{b}$ of each halo with mass $M_{\rm h}$ 
that collapsed at $z$, denoted $f_\mathrm{b} (M_{\rm h}, z)$, is 
calculated by equation~(\ref{eq:fb-O08}) using a fitting formula for
$M_\mathrm{c}(z)$:
\begin{eqnarray}
\label{eq:fiteq}
M_{\rm c}(z) &=& 6.5\times 10^9 \nonumber \\
&&\times \exp(-0.604z)  \exp{[-(z/8.37)^{17.6}]}~h^{-1}~\Msun.
\end{eqnarray}
This fitting formula is derived from the result of simulation of 
\cite{Okamoto2008} in which the reionization occurs at $z = 9.0$.
Figure \ref{fig:Mcz} shows $M_{\rm c}$ as a function of redshift.
While the minimum halo mass of the lower resolution models 
($M_{\rm min} = 8.79\times10^9~h^{-1}~\Msun$) is larger 
than $M_{\rm c}$, those of the higher resolution models 
($M_{\rm min} = 1.10\times10^9~h^{-1}~\Msun$ for the $\nu^2$GC-H1 
and $M_{\rm min} = 1.37\times10^8~h^{-1}~\Msun$ for the $\nu^2$GC-H2, -H3,
respectively) are smaller than $M_{\rm c}$ at low redshift.
In the halos with mass below $M_{\rm c}$, the gas heating by
the UV background affects the properties of galaxies.
To mimic this effect, \cite{Somerville2002} also adopt the
formulation of \cite{Gnedin2000}; however, they assume that
$M_{\rm c}(z)$ is given by constant $V_{\rm circ} = 50~{\rm km}~{\rm s^{-1}}$.
In figure \ref{fig:Mcz}, we also show the redshift evolution of the halo 
mass with the fixed circular velocity of $V_{\rm circ} = 17$, 
$30$ and $50~{\rm km}~{\rm s}^{-1}$, respectively.
We can see that the behavior of the $M_{\rm c}$ proposed by \cite{Okamoto2008}
and the $M_{\rm c}$ given by fixed $V_{\rm circ}$ are significantly
different.

In this paper we treat the effect of the gas heating due to the
cosmic UV background as follows.
For a halo with total baryonic mass (i.e.,
the sum of the masses of the stars, SMBH, cold gas, and hot gas of all
galaxies in the halo) of $M_\mathrm{b,tot} \le f_\mathrm{b} (M_{\rm 
h}, z) M$, the baryonic mass of $f_\mathrm{b} (M_{\rm h}, z) M_{\rm h} - M_\mathrm{b, tot}$ is
added to the halo as hot gas with temperature of $T_\mathrm{vir}$.  On
the other hand, for the halos with $M_\mathrm{b, tot} > f_\mathrm{b}
(M_{\rm h}, z) M_{\rm h}$, an appropriate mass of hot gas is removed from the halo
keeping its metallicity unchanged so that the mass fraction of baryon
in the halo coincides with $f_\mathrm{b} (M_{\rm h}, z)$; this prescription
mimics photoevaporation by UV background radiation during the
reionization. 
When $M_\mathrm{b, tot} - M_\mathrm{hot} > f_\mathrm{b}(M_{\rm h}, z) M_{\rm h}$, 
we have to reduce the additional cold gas masses in order to 
the mass fraction of baryon in the halo coincides with $f_\mathrm{b} (M_{\rm h}, z)$. 
However cold gas is much denser than hot gas, and may be self-shielded
from the UV background radiation.
Therefore we assumed that the cold gas component is not affected by
the UV heating and allowed such halos to have larger baryon mass 
than $f_{\rm b}(M_{\rm h},z)M_{\rm h}$.
Note that the fraction of such halos is less than $1 \%$, thus
the treatment of the cold gas in this process does not have significant
effects on the results presented in this paper.

Although \cite{Okamoto2008} found that such
photoevaporation is important particularly just after the reionization
(see middle panel of figure~5 of \citealt{Okamoto2008}), the effect in
our model is assumed to occur for such less-massive halos with
$M_\mathrm{b, tot} > f_\mathrm{b} (M_{\rm h}, z) M_{\rm h}$, regardless of their
collapsing redshifts when $z \le z_\mathrm{reion}$.

\subsection{Star formation and feedback in disk}\label{sec:sffb}
In this section we describe the star formation in cold gas disk and 
the reheating of cold gas by SNe.
Our implementation follows the standard recipe adopted in other SA models
(e.g., \citealt{Cole2000}).

The cooling process of diffuse hot gas is followed by star formation in the
cold gas disk. The star formation rate (SFR), $\psi$, is given by $\psi=M_{\rm cold}/
\tau_{\rm star}$, where $M_{\rm cold}$ is the cold gas mass,
and $\tau_{\rm star}$ is the time scale of the star formation (SF).
We assume that the star formation activity in galaxy disk 
is related to the dynamical time scale of the disk, $\tau_{\rm d} \equiv
 r_{\rm d} / V_{\rm d}$, where $r_{\rm d}$ and $V_{\rm d}$ are
the disk radius and disk rotation velocity, respectively, 
defined in subsection \ref{sec:response}.
Thus we adopt the following formula for star formation time scale $\tau_{\rm star}$,
\begin{equation}
 \tau_{\rm star}=\varepsilon_{\rm star}^{-1}\tau_{\rm d}\left[1+\left(\frac{V_{d}}{V_{\rm
 hot}}\right)^{\alpha_{\rm star}}\right],
\end{equation}
where $\varepsilon_{\rm star}$, $\alpha_{\rm star}$, and $V_{\rm hot}$ are free parameters.
Although the above modeling of the SF time scale well reproduces the several
physical properties of observed galaxies as shown later (see section \ref{sec:localresults}),
there could be other modeling for the SF time scale.
For example, we have examined the model in which the SF time scale depends on 
the global dust surface density of galaxy,
and found that the choice of SF time scale model could have significant 
effect on the galaxy formation history \citep{Makiya2014}.
Although this would be promising for reproducing many aspects of observed galaxies, 
in this paper, we adopt this kind of standard prescription of star formation for simplicity.

Consequent to supernova explosion, we assume that a fraction of the cold 
gas is reheated and ejected from galaxy at a rate of ${M_{\rm cold}}/{\tau_{\rm reheat}}$, where the time scale of
reheating, $\tau_{\rm reheat}$ is given as follows:
\begin{equation}
\tau_{\rm reheat}=\frac{ \tau_{\rm star}}{\beta(V_{d})},
\end{equation}
with
\begin{equation}
\beta(V_{d})\equiv\left(\frac{V_{d}}{V_{\rm hot}}
\right)^{-\alpha_{\rm hot}},
\label{eqn:vhot}
\label{eqn:beta}
\end{equation}
where $V_{\rm hot}$ and $\alpha_{\rm hot}$ are free parameters.

With the above equations, we obtain the masses of hot
gas, cold gas, and disk stars as functions of time (or redshift).  
The chemical enrichment associated with star formation and SN feedback
is treated by extending the work of \citet{Maeder1992}.  For
simplicity, instantaneous recycling is assumed for SNe II, and any
contribution from SNe Ia is neglected.

In summary, the baryon evolution during the star formation process
is described by the following equations:
\begin{eqnarray}
 \dot{M}_{\rm star}&=&\alpha\psi(t), \label{eq:baryonevo1}\\
 \dot{M}_{\rm hot}&=& \beta\psi(t),\\
 \dot{M}_{\rm BH}&=& f_{\rm BH}\psi(t),\\
 \dot{M}_{\rm cold}&=&-(\alpha+\beta+f_{\rm BH})\psi(t),\\
 (M_{\rm cold}Z_{\rm cold})\dot{}&=&[p-(\alpha+\beta+f_{\rm BH})Z_{\rm cold}]\psi,\label{eqn:p}\\
 (M_{\rm hot}Z_{\rm hot})\dot{}&=&\beta Z_{\rm cold}\psi, \label{eq:baryonevo5}
\end{eqnarray}
where $M_{\rm star}$ and $M_{\rm hot}$ are the masses of stars and hot gas, 
respectively; $\psi={M_{\rm cold}}/{\tau_{\rm star}}$ is SFR, 
$Z_{\rm cold}$ and $Z_{\rm hot}$ are the 
metallicities of cold and hot gases, respectively,
and $M_{\rm BH}$ is the mass of the nuclear SMBH.
The constant parameter $f_{\rm BH}$ controls the 
accretion rate of cold gas onto the SMBH during the starburst.
In ordinary star formation process in disk,
we assume that no cold gas gets accreted by the SMBH (i.e., $f_{\rm BH} = 0.0$).
The galaxy merger and SMBH evolution will be detailed 
in later subsections (\ref{sec:merger} and \ref{sec:SMBH}).
The locked-up mass fraction $\alpha$ and chemical yield 
$p$ are chosen to be consistent with initial mass function (IMF).
For the Chabrier IMF 
(\citealt{Chabrier2003}), which is adopted in our standard model,
$\alpha = 0.52$ and $p = 1.68$ $Z_{\odot}$.

We can solve these equations analytically as
\begin{eqnarray}
 \Delta M_{\rm cold}(t)&=&M_{\rm cold}^{0}\left\{1-\exp\left[-(\alpha+\beta)\frac{t}{\tau_{\rm star}}\right]\right\},\\
 \Delta M_{\rm star}(t)&=&\frac{\alpha}{\alpha+\beta}\Delta M_{\rm cold}(t),\\
 \Delta M_{\rm hot}(t)&=&\frac{\beta}{\alpha+\beta}\Delta M_{\rm cold}(t),\\
 Z_{\rm cold}(t)&=&Z_{\rm cold}^{0}+p\frac{t}{\tau_{\rm star}},\\
 Z_{\rm hot}(t)&=&\left[M_{\rm hot}^{0}Z_{\rm
 hot}^{0}+\frac{\beta}{\alpha+\beta}\left\{\frac{}{}\right.\right.\nonumber\\
 &&\left(\frac{p}{\alpha+\beta}+Z_{\rm
 cold}(t)\right)\Delta M_{\rm cold}(t) \nonumber\\
 &&-(Z_{\rm cold}(t)-Z_{\rm
 cold}^{0})M_{\rm cold}^{0}{\left.\left.\frac{}{}\right\}\right]}/M_{\rm hot}(t),
\
\end{eqnarray}
where the $\Delta$ symbol indicates that the variable is incremented or decremented 
in the current time step.
All $\Delta$ variables are defined as positive. 
The superscript 0 denotes an initial values at the beginning of the time-step
(i.e., $t = 0$).
Note that here we assumed $f_{\rm BH} = 0$.
For the case of burst-like star formation induced by major merger,
see subsection~\ref{sec:merger}.

\subsection{Mergers of galaxies and formation of spheroids}\label{sec:merger}
After the merging of DM halos, the newly formed halo should contain two or more
galaxies.
The central galaxy in the most massive progenitor halo is designated as 
the central galaxy of newly formed halo, while the others are regarded as satellite galaxies.
These satellite galaxies will fall into the central galaxy by dynamical friction
 (central-satellite merger).
We set the merger time scale due to the dynamical friction as $\tau_{\rm mrg}=f_{\rm mrg}\tau_{\rm fric}$,
where $f_{\rm mrg} \sim 1$ is adjustable parameter and $\tau_{\rm fric}$ is
the time scale of dynamical friction.
If $\tau_{\rm mrg}$ is shorter than the time elapsed since the satellite
galaxy enters the common halo, the satellite and central galaxy are merged.
We reset this elapsed time to zero when the host halo mass doubles.

For the time scale of dynamical friction, $\tau_{\rm fric}$, we adopt the
formulation of \cite{Jiang2008, Jiang2010} which is obtained by fitting to the 
cosmological $N$-body simulations :
\begin{equation}
\tau_{\rm fric} = \frac{f(\varepsilon)}{2C}\frac{V_{\rm circ}R_{\rm circ}^2}
{GM_{\rm s}\ln{(1+M_{\rm h}/M_{\rm s})}}
\end{equation}
where $C = 0.43$ is a constant fitting parameter, 
$V_{\rm circ}$ is the circular velocity of the common halo, 
$R_{\rm circ}$ is the radius of the circular orbit of the satellite 
halo, and $M_{\rm h}$ and $M_{\rm s}$ are the total mass of the host 
halo and satellite halo, respectively.
We simply assumed that $R_{\rm circ} = R_{\rm h}$ where $R_{\rm h}$ is 
the virial radius of the host halo.
The function $f(\varepsilon) = 0.90\varepsilon^{0.47}+0.60$ accounts 
for the dependence of $\tau_{\rm fric}$ on the orbital circularity $\varepsilon$.
We set to $\varepsilon = 0.5$, which is the average value of $\varepsilon$ 
estimated by high-resolution $N$-body simulations performed by \cite{Wetzel2011}.
In our $N$-body simulations we can resolve the satellite halos even after they
entered the common halo, and therefore the time scale of dynamical 
friction can be directly drown from simulations; however, in this paper 
we adopted simplified formula described above to save the computational time. 
The effect of this simplification will be examined in a future paper.

The satellite galaxies can randomly collide and merge (satellite-satellite merger).
\citet{Makino1997} conducted an {\it N}-body simulation of a system of same mass
galaxies.
They find that a merger rate, $k_{\rm MH}$, is described by following simple scaling
in this situation:
\begin{eqnarray}
 k_{\rm MH}&=&\frac{N}{500}\left(\frac{1~\rm Mpc}{R_{\rm h}}\right)^{3}
  \left(\frac{r_{\rm gal}}{0.1~\rm Mpc}\right)^{2}\nonumber\\
  &&\times\left(\frac{\sigma_{\rm gal}}{100~\rm km~s^{-1}}\right)^{4}
  \left(\frac{300~\rm km~s^{-1}}{\sigma_{\rm halo}}\right)^{3} \mbox{Gyr}^{-1},
\end{eqnarray}
where $N$, $\sigma_{\rm gal}$, $r_{\rm gal}$ and $\sigma_{\rm halo}$ are 
the number of satellite galaxies,
one-dimensional velocity dispersions of the galaxy, galaxy radius and parent halo, respectively.  
In our model a satellite galaxy will collide with another satellite galaxy picked out
at random, with the probability
of $\Delta t \times k_{\rm MH}$, where $\Delta t$ is the time step of the calculation.

We consider two distinct modes for galaxy merger, i.e., major merger and minor merger.
If the ratio of 
baryonic mass (stars, cold gas, hot gas and SMBH mass) 
of two merging galaxies, $f$ ($< 1$), exceeds the critical value 
$f_{\rm bulge}$, major merger occurs.
Major mergers induce burst-like star formations, in which all of 
the cold gas in the merging system turns into stars and hot gas.
The star formation and SN feedback law is the same with the disk star 
formation (see section~\ref{sec:sffb}), except for assuming the very 
short star formation time scale ($\tau_{\rm star} \rightarrow 0$).
The bulges and stellar disks of the progenitor completely reform
into the bulge component of the new galaxy,
together with the stars born during the merger.
Note that when applying the SN feedback law, the disk velocity $V_{d}$ 
is replaced by the velocity dispersion of the new bulge $V_{b}$ 
(defined in subsection~\ref{sec:response}).

On the other hand, if $f<f_{\rm bulge}$, a minor merger occurs.
In this case, stellar and cold gas components of 
the smaller galaxy are absorbed into the bulge and cold gas 
disk of the larger galaxy, respectively, with no starburst events.

\subsection{Supermassive black holes}
\label{sec:SMBH}
Along with the evolution of galaxies, SMBHs at galaxy centers also evolve
by the following mechanismas: 
(1) SMBH coalescence, (2) accretion of cold gas (during a major merger of galaxies),
and (3) ``radio-mode'' gas accretion.
Note that we assume a central SMBH in every galaxy.
When the galaxies first collapsed, the seed BH have formed with 
mass $M_{\rm seed}$, which is a tunable parameter.

It has been shown by theoretical studies that a major merger of galaxies 
can drive substantial gaseous inflows into a galaxy center
\citep[e.g.,][]{Mihos1994, Mihos1996, Barnes96, DiMatteo05, Hopkins05, Hopkins06}.
We assume that a fraction of this inflowing cold gas gets accreted by the central SMBH.
The mass of cold gas accreted by the SMBH, $\Delta M_{\rm BH}$ is modeled as follows:
\begin{eqnarray} 
\Delta M_{\rm BH} &=& f_{\rm BH} \Delta M_{\rm star, burst}, \label{eq:bhaccret}
\end{eqnarray} 
where $f_{\rm BH}$ is a constant, and $\Delta M_{\rm star, burst} $ is the
total mass of stars formed during the starburst. We set $f_{\rm BH} = 0.005$ to
 match the observed relation between masses of host bulges and
SMBHs at $z = 0$ (see subsection \ref{sec:SMBHobs}).
The accretion of cold gas triggers the quasar activities.
For more detailed model of quasars, see \cite{Enoki2003}, \cite{Enoki2014},
and \cite{Shirakata2015}.

Considering the very short time scale of starburst ($t/\tau_{\rm star} \rightarrow \infty$) 
assumed here and the mass accretion onto the nuclear SMBH,
we solve equations~(\ref{eq:baryonevo1})--(\ref{eq:baryonevo5}) to obtain the following:
\begin{eqnarray}
 \Delta M_{\rm star}   &=& \frac{\alpha}{\alpha+\beta+f_{\rm BH}}M_{\rm cold}^{\rm 0},\\
 \Delta M_{\rm hot} &=& \frac{\beta}{\alpha+\beta+f_{\rm BH}}M_{\rm cold}^{\rm 0},\\
 \Delta M_{\rm BH} &=& \frac{f_{\rm BH}}{\alpha+\beta+f_{\rm BH}}M_{\rm cold}^{\rm 0},\\
 \Delta (M_{\rm hot}Z_{\rm hot})&=&\frac{\beta}{\alpha+\beta+f_{\rm BH}} \nonumber \\
 &\times & \left(\frac{p}{\alpha+\beta+f_{\rm BH}}+Z_{\rm cold}^{0} \right)M_{\rm cold}^{0},
\end{eqnarray}
where $\Delta M_{\rm star}$, $\Delta M_{\rm hot}$, $\Delta M_{\rm BH}$, and 
$\Delta (M_{\rm hot} Z_{\rm hot})$ are the increasing amount of the stellar mass,
hot gas mass, BH mass and the metal mass in the hot gas, respectively, 
during a starburst.
The superscript 0 indicates the total values in the merger progenitors. 
We again emphasize that all the cold gas is exhausted in our starburst model.

During a merger event, an SMBH also increases its mass 
via the SMBH--SMBH coalescence.
In this paper, we simply assume that SMBHs merge instantaneously right
after the merger of their host galaxies, because it is difficult to
estimate the time scale of SMBH mergers owing to the existence of many
complicated physical processes such as the dynamical friction, 
stellar distribution, multiple SMBH interaction, and gas dynamical
effects (see, e.g., \citealt{Colpi2014}).  
As shown in \cite{Enoki2004}, the mass growth of SMBHs in our model
is mainly due to the gas accretion during major merger, at least, at z
$\lesssim$ 1, and therefore therefore the assumption of
the instantaneous coalescence does not have significant effects.
The other evolution channel, radio-mode gas accretion 
related to the AGN feedback process, is described next.

\subsection{AGN feedback}
To reproduce the observed break at the bright end of the luminosity
functions (LFs), 
we introduced the so-called radio-mode AGN feedback process into our model.
In this mode, the hot gas accreted by SMBH powers radio jet that    
injects energy into the hot halo gas, quenching the cooling of the hot gas 
and resultant star formation in the massive halo.
Radio-mode AGN feedback is also expected to contribute to
the downsizing evolution of galaxies \citep[e.g.,][]{Croton2006, Bower2006}.

Our implementation of AGN feedback follows the formulation of \cite{Bower2006}.
In their formulation, gas cooling in the halo is inhibited
when the following conditions are satisfied:
\begin{equation}
\alpha_{\rm cool} t_{\rm dyn}(r_{\rm cool}) < t_{\rm cool}
\label{eq:AGNt}
\end{equation}
and
\begin{equation}
\varepsilon_{\rm SMBH} L_{\rm Edd} > L_{\rm cool},
\label{eq:AGNe}
\end{equation}
where $t_{\rm dyn}$ is the dynamical time scale of the halo at the cooling radius, 
$t_{\rm cool}$ is the time scale of gas cooling, $L_{\rm Edd}$ is the
Eddington luminosity of the AGN, $L_{\rm cool}$ is the cooling
luminosity of the gas, and $\alpha_{\rm cool}$ and $\varepsilon_{\rm
  SMBH}$ are free parameters that are tuned to reproduce the
observations.
Under these conditions, AGN feedback is limited to haloes in quasi-hydrostatic 
equilibrium, and having a sufficiently evolved SMBH.
In the halo experiencing AGN feedback, the SMBH at the center grows
by accreting hot halo gas.
\cite{Bower2006} assumed that the accretion flow is automatically adjusted
by itself so that heating luminosity balances with cooling luminosity, namely,
the accretion rate is set to $\dot{M}_{\rm BH} = L_{\rm cool}/\eta c^2$.
Here $\eta$ is the radiative efficiency. We assumed $\eta = 0.1$ for all SMBHs, 
based on the observational estimation of \cite{Davis2011}.
The value of $\eta$ does not significantly affect on the results
since the mass growth of SMBHs is dominated by the gas accretion during 
major merger.

\subsection{Size of galaxies and dynamical response to gas removal}
\label{sec:response}
This subsection explains how we estimate the galaxy size, 
disk rotation velocity and velocity dispersion of bulges.
Our recipe of size estimation almost follows the procedure of \cite{Cole2000}.

\subsubsection{Disk formation from cooled gas}
First, we estimate the size of galaxy disk as follows.
We assume that the hot halo gas has the same specific angular momentum
as the DM halo and collapses to the cold gas disk 
while conserving the angular momentum.
We introduce the dimensionless spin parameter $\lambda_{\rm H}$ 
as $\lambda_{\rm H}\equiv L|E|^{1/2}/GM^{5/2}$,
where $L$ is the angular momentum, $E$ is the binding energy
and $M$ is the DM halo mass.
Although the distribution of $\lambda_{\rm H}$ is often approximated 
by a log-normal distribution (e.g., \citealt{mmw98}),
it has been known that the distribution of $\lambda_{\rm H}$
deviates from log-normal in large $N$-body simulations (e.g., \citealt{Bett2007}).
However, according to \cite{Bett2007}, the shape of distribution 
depends on halo finding algorithm and the log-normal function is still
slightly better for FoF halos than the modified fitting function proposed by them.
In this paper, we simply adopt the log-normal distribution 
\begin{equation}
p(\lambda_{\rm H})d\lambda_{\rm H}=
\frac{1}{\sqrt{2\pi}\sigma_{\lambda}}
\exp\left[-\frac{(\ln\lambda_{\rm H}-\ln\bar{\lambda})^2}
{2\sigma_{\lambda}^{2}}\right] d\ln\lambda_{\rm H},\label{eqn:lognormal}
\end{equation}
where $\bar{\lambda}$ and $\sigma_{\lambda}$ denote the mean
and logarithmic variance of the spin parameter, respectively.
Here we use $\bar{\lambda}=0.042$ and $\sigma_{\lambda}=0.26$., which are the value 
obtained by \citealt{Bett2007} for FoF halos.

Using the spin parameter $\lambda_{\rm H}$,
the effective radius $r_{d}$ of a resultant cold gas disk is expressed as 
follows \citep{f79, fe80, f83}:
\begin{equation}
r_{d}=(1.68/\sqrt{2})\lambda_{\rm H}R_{i},
\label{eq:rdisk}
\end{equation}
where the initial radius of the hot gas sphere, $R_{i}$, is set to 
the virial radius of the host halo or the cooling radius, whichever is smaller.
In each time step, the disk size of central galaxies are updated 
if their disk mass have increased from the previous time step.
At this time, we set the disk rotation velocity $V_{d}$ 
to be the circular velocity of its host halo.

\subsubsection{Dynamical response to disk star formation}
After the formation of rotationally supported disks, the SN feedback
subsequent to disk star formation expels cold gas continuously.
As the baryonic mass of galaxies decreases, the gravitational
potential well becomes shallower, depending on the mass ratio of baryons to
DM within the galactic disk.  In response to the variation of
the depth of the gravitational potential well, gravitationally bound
systems expand and their rotation speed slows down \citep{ya87}.  We
refer to this effect as the {\it dynamical response} here.  
Dwarf galaxies having
shallow gravitational potential wells and therefore suffered significant
SN feedback are affected more by the dynamical response.  Using our SA
models taking into account this for starburst, we have shown that this
affects the scaling relations of elliptical galaxies especially for
dwarfs \citep{Nagashima2004, Nagashima2005}. See those papers for the scheme of
introducing the dynamical response in SA models.
In the present paper, we also apply
this effect to disk evolution.  

The basic result for disks used here is given by \citet{Koyama2008}.  
At first, we assume a galactic disk within a static DM halo and
approximate the density distributions of disks and DM as the
Kuzmin disk \citep{kuzmin1952, kuzmin1956} and the Navarro-Frenk-White
(NFW) profile \citep{nfw97}, respectively.  Then, we consider that the
gas mass of disks gradually decreases due to the SN feedback, that is,
the so-called adiabatic mass-loss, and that the dynamical response to
the gas removal on the disks.
The initial radius of cold gas disk is calculated by equation~(\ref{eq:rdisk}).
Here we assume that the disk size is determined only by the
gravitational potential of the host dark halo, conserving the specific
angular momentum of the cooled gas.  Although this might be too simple
because the central region of galaxies would form dynamically with
cooling gas, it should be a good approximation for the outer disk.
Thus we take this treatment in this paper as usual.

Here we define ${\cal M}$ and $R$ as ratios of mass and size at a final
state relative to those at an initial state, and $z_i$ and $z_f$ as
ratios of baryonic disk size relative to size of dark halos at those
states.  According to \citet{Koyama2008}, we obtain
\begin{equation}
{\cal M} = \frac{1}{R} + \frac{q(z_f) - z_i q(z_i) / z_f}{m_i},
\end{equation}
where $m_i$ is the mass ratio of baryons to dark matter at the initial
state, and $q(z)$ is a function depending on the distributions of
baryons and dark matter.  In this case, we cannot obtain an analytic
form of $q(z)$.  Instead, we expand the above equation around $z=0$ and
$R-1\simeq 0$ as
\begin{equation}
{\cal M} = \frac{1}{R} + D (R-1),
\end{equation}
where
\begin{eqnarray}
D  =  && \frac{c}{m_i} \left[\ln (1+c) - \frac{c}{1+c}\right]^{-1} 
 \left[ c z_i^2 \left( 3 + 2\ln\frac{cz_i}{2}\right) \right. \nonumber\\
&&  \qquad\left. - \frac{16}{3}c^2 z_i^3
 - c^3 z_i^4 \left( \frac{33}{8} + \frac{9}{2}\ln\frac{cz_i}{2}\right)\right],
\end{eqnarray}
and $c$ is the concentration parameter described in section~\ref{sec:Gascooling}. 
Note that we take a higher order term of
$z_i$ for $q(z)$ than that written in equation~(A5) in
\citet{Koyama2008}.

The approximation used here is justified as follows.  It is expected
that the change in sizes and disk rotation velocities during a time-step
is very small because of the quiescent star formation, and that the size
of baryonic disks is smaller sufficiently than that of dark halos.
These mean $R-1 \ll 1$ and $z_i \ll 1$.
We have checked that these assumptions are indeed validated in our model.

The change of the disk rotation velocity is given by
\begin{equation}
U\equiv\frac{V_{d,f}}{V_{d,i}} = \left[ \frac{m_f/z_f + 4f(z_f)}{m_i/z_i + 4f(z_i)}\right]^{1/2},
\end{equation}
where $f(z)$ is also a function depending on the distributions of baryons and dark matter, 
similar to $q(z)$.  The form of $f(z)$ is shown in equation~(A4) in \citet{Koyama2008}.

We would like to recall here the well-known results for non-dark matter
case, $R=V^{-1}={\cal M}^{-1}$.  These relations are obtained by setting
$f(z)$ and $q(z)$ to zero.  In the opposite limiting case, because dark
matter dominates, $q(z)$ becomes much larger.  In this case, even if
${\cal M}$ becomes zero, $R$ and $U$ do not vary.  This corresponds to
the case discussed in \citet{DekelSilk1986}.

The effect of the dynamical response on the disk shall be discussed in
detail in another paper.

\subsubsection{Dynamical Response to starburst and spheroidal remnants}
The size of the bulge formed in a major merger is characterized
by the virial radius of the baryonic component.  Applying the virial
theorem the total energy in each galaxy is calculated as follows:
\begin{equation}
E_{i}=-\frac{1}{2}[
(M_{b,i}+M_{{\rm BH},i})V_{b,i}^{2}+(M_{d,i}+M_{\rm cold})V_{d,i}^{2}],
\end{equation}
where $M_{b}$, $M_{\rm BH}$, $M_{d}$ and $M_{\rm cold}$ are the masses 
of the bulge, central black hole, stellar disk and cold gas disk, 
respectively, and $V_{b}$ and $V_{d}$ denote the velocity dispersion 
of the bulge and the rotation velocity of the disk, respectively.
The subscript $i = \{0,1,2\}$ indicate the merged galaxy, 
larger progenitor galaxy and smaller progenitor galaxy, 
respectively.
Furthermore the orbital energy $E_{\rm orb}$ between 
the progenitors just before the merger is given as follows:
\begin{equation}
E_{\rm orb} = - \frac{E_1 E_2}{(M_2/M_1)E_1+(M_1/M_2)E_2}.
\end{equation}
By energy conservation, we obtain the following:
\begin{equation}
\label{eqn:fdiss}
f_{\rm diss}(E_{1}+E_{2}+E_{\rm orb})=E_{0},
\end{equation}
where $f_{\rm diss}$ is the fraction of energy dissipated from
the system during major merger.
The rate of energy dissipation depends on
complicated physical processes such as the viscosity and friction due to gas. 
In this paper, we simply parameterize $f_{\rm diss}$ as follows:
\begin{equation}
f_{\rm diss} = 1+\kappa_{\rm diss}f_{\rm gas},
\end{equation}
where 
\[
f_{\rm gas} = \frac{M_{\rm cold}}{M_{\rm star}+M_{\rm cold}+M_{\rm BH}}
\]
is the gas mass fraction of the merging system and $\kappa_{\rm diss}$
is a dimensionless parameter.
Here we set $\kappa_{\rm diss}=2.0$ to reproduce
the distribution of size and velocity dispersion of elliptical galaxies
(see section \ref{sec:Esize}).
There are several studies on this issue by using hydrodynamical simulations
and SA models, and it is confirmed that above parameterization of
$f_{\rm diss}$ can be a good approximation 
(see, e.g., \citealt{Hopkins2009}; \citealt{Shankar2013}).

We assumed that there remains only the bulge component supported by 
velocity dispersion just after the merger.
Therefore the velocity dispersion and the size of merger remnant 
can be estimated from following equations,
\begin{equation}
E_{\rm 0}
= -\frac{1}{2}M_{{\rm tot}, 0}V_{b,0}^2,
\end{equation}
and
\begin{equation}
r_{{\rm b},0}=\frac{GM_{\rm tot,0}}{2V_{b,0}^{2}},\label{eqn:sizeE}
\end{equation}
where $M_{{\rm tot},0}$ is the total baryonic mass of the merger remnant.

\if
(以下, dynamical responce について. あとに回す.)
In the case of the formation of elliptical galaxies driven by major
mergers of galaxies, the size and velocity dispersion are estimated,
explicitly taking into account the dynamical response to
starburst-induced gas removal.  The outline is as follows: At first, two
or more galaxies merge.  During the merger, a fraction $f_{\rm diss}$ of
the total energy of baryonic matter is assumed to be dissipated and a
spheroidal merger remnant is formed and reaches the virial equilibrium
immediately.  Subsequently, the starburst occurs and a fraction of gas is
gradually expelled from the spheroidal system according to the SN
feedback law.  Because the gravitational potential varies during the gas
removal, the system expands and its velocity dispersion is lowered.  The detail is shown below.
\fi

As a consequence of star formation and SN feedback, part of gas removed
from galaxies and the mass of the system changes.  At this time the
structural parameters of galaxies also changes due to the dynamical
response.  We include this effect into our model adopting the Jaffe
model \citep{jaffe}.  In this paper we assume the case of slow
(adiabatic) gas removal compared with the dynamical time scale of the
system, similar to that for disks.  For the case of rapid gas removal,
we refer the reader to \cite{Nagashima2003}.  According to the
\cite{Nagashima2003}, the effect of dynamical response becomes stronger
in the case of non-adiabatic gas removal.  Therefore the assumption of
the adiabatic gas removal should be considered as conservative.  We
should keep in mind that the effect might be stronger for dwarf
ellipticals having shorter time scale of gas removal compared to giants.

Defining by ${\cal M}, R$ and $U$ the ratios of mass, size and velocity
dispersion at a final state relative to those at an initial state, the
response under the above assumption is approximately given by
\begin{eqnarray}
R&\equiv&\frac{r_{f}}{r_{i}}=\frac{1+D/2}{{\cal M}+D/2},\label{eqn:sizeE2}\\
U&\equiv&\frac{V_{b,f}}{V_{b,i}}=\sqrt{\frac{{\cal M}/R+Df(z_{f})/2}{1+Df(z_{i})/2}},
\end{eqnarray}
where $D=1/y_{i}z_{i}^{2}$, $y$ and $z$ are the ratios of density and
size of baryonic matter to those of dark matter.  We use equation~(36)
in \citet{Nagashima2005} for the form of $f(z)$.  The subscripts $i$ and
$f$ stand for the initial and final states in the mass loss process.
Note that $U$ is the ratio of velocity dispersion, different from that
for disks.  The contribution of dark matter is estimated from the
central circular velocity of halos, $V_{\rm cent}$, which is defined
below.

\subsubsection{Back reaction of dynamical response to dark halos}
When galaxies suffer the dynamical response to gas removal caused by the
SN feedback, the dark matter within a central region of dark halos
hosting the galaxies must also suffer the dynamical response as its back
reaction.  
For simplicity we compute the dynamical response on the dark matter 
distribution after the computation of the dynamical response on baryons,
although they occur simultaneously in reality.
Here we ignore the effect of the so-called adiabatic contraction for
dark matter during the condensation of cooled gas.  This is because
the central region of galaxies should form not adiabatically but
dynamically.  Thus we assume that the cooled gas condenses and relaxes
dynamically together with dark matter and is removed adiabatically by
the SN feedback affecting the central region of dark halos as the back
reaction.  This would require detailed research by using hydrodynamic
numerical simulations.

Here we focus on the region within the half-mass radius of central galaxies, at 
which the density of baryons is expected to be comparable to that of dark matter.  
To take into account this process, we define a central circular velocity of dark
halo $V_{\rm cent}$, approximately within the effective radius of the
central galaxy.  When a dark halo collapses without any progenitors,
$V_{\rm cent}$ is set to $V_{\rm circ}$.  After that, although the mass
of the dark halo grows by subsequent accretion and/or mergers, $V_{\rm
cent}$ remains constant or decreases with the dynamical response.  When
the mass is doubled, $V_{\rm cent}$ is set to $V_{\rm circ}$ again.
According to \citet{Nagashima2004} and \citet{Nagashima2005}, we assume
that $V_{\rm cent}$ is lowered by the dynamical response to mass loss
from a central galaxy of a dark halo by SN feedback as follows,
\begin{equation}
\frac{V_{{\rm cent},f}}{V_{{\rm cent},i}}=
 \frac{M_{f}/2+M_{d}(r_{i}/r_{d})}{M_{i}/2+M_{d}(r_{i}/r_{d})}.
\end{equation}
The change of $V_{\rm cent}$ in each time-step is only a few per cent.
Under these conditions, the approximation of static gravitational
potential of dark matter is valid even during starbursts.  This also
applies to subhalos.  Rigorously speaking, we must assume an isothermal
distribution of dark matter, in which the density is proportional to the
inverse of $r^2$, because the above equation indicates the dynamical
response to mass loss within the half-mass radius of central galaxies.
In spite of this, this should be good approximation because the NFW
profile has a slope $-1$ and $-3$ within and outside the core radius,
respectively, which means that we can expect that the effective slope
would be approximately $-2$.
Of course this expectation is optimistic since we considered the inner 
region of a halo where a slope is $-1$. However, we need detailed 
hydrodynamical simulations to know the actual mass profile since
the adiabatic contraction due to gas cooling would affect the slope.

Once a dark halo falls into its host dark halo, it is regarded as a
subhalo.  Because subhalos do not grow in mass in our model, the central
circular velocity of the subhalos monotonically decreases.  Although the
change of $V_{\rm cent}$ during a time-step is small, accumulated change
cannot be negligible owing to the monotonicity.  Therefore, this affects
the time scales of mergers.

The details of the dynamical response are shown in \citet{Nagashima2003,
Nagashima2004} and \citet{Nagashima2005} for bulges and
\citet{Koyama2008} for disks.  The effect of the dynamical response is
the most prominent for dwarf galaxies of low circular velocity because
of the substantial removal of gas due to strong SN feedback \citep{ya87,
Nagashima2004}.  If the dynamical response had not been taken into
account, velocity dispersions of dwarf ellipticals would have been much
larger than those of observations, determined only by circular
velocities of small dark halos in which dwarf ellipticals resided.  For
giant ellipticals, on the other hand, the effect of the dynamical
response is negligible because only a small fraction of gas can be
expelled due to weak SN feedback.  Similarly, for disks, in order to
reproduce the observed Tully-Fisher relation, the dynamical response on
disks is required.  Otherwise, the slope for dwarf spirals becomes
different from observed one as shown in \citet{Nagashima2005}.  This
point will be discussed in detail in another paper.

\subsection{Photometric properties and morphological identification}
\label{sec:photo}
Calculating the baryonic processes described in the above subsections,
we finally obtain the SF hand metal enrichment histories of each 
galaxies.
From this information, we can calculate spectral energy distribution (SED)
of model galaxies by using a stellar population synthesis code 
of \cite{Bruzual2003}.

To estimate the extinction of starlight, 
we first assume that the dust-to-cold gas ratio is proportional to 
the metallicity of the cold gas; second, we assume that the dust 
optical depth is proportional to the dust column density. 
The dust optical depth $\tau_{\rm dust}$ is then calculated as follows:
$\tau_{\rm dust}$ is given by
\begin{equation}
 \tau_{\rm dust} = \tau_{0} \left(\frac{M_{\rm cold}}{\Msun}\right)
 \left(\frac{Z_{\rm cold}}{Z_{\odot}}\right)
 \left(\frac{r_{\rm d}}{\rm kpc}\right)^{-2}
\label{eqn:dust}
\end{equation}
where $r_{\rm d}$ is the effective radius of the galaxy disk and $\tau_{0}$ is a
tunable parameter that should be chosen to fit the local observations (such
as LFs).
The wavelength dependence of optical depth is assumed to follow 
Calzetti-law \citep{Calzetti2000}.  
Dust distribution is assumed to obey the slab dust model \citep{ddp89} for disks.

In our model, a major merger induces starburst activity, in which
all the cold gas turns into stars and hot gas.
Therefore, no cold gas and dust exist immediately after the starburst.
Hence, the dust optical depth exactly equals to zero and
galaxy color becomes too blue compared to the observations.
To avoid this problem, we estimate the amount of dust extinction 
during the starburst as follows.
First, we randomly assign the merger epoch within the current time step. 
Second, we calculate the amount of remaining dust at the end of the time step.
At this time, the time scale of gas consumption during the burst is assumed 
as the dynamical time scale of the merged system, $r_{\rm b}/V_{ \rm b}$.
The dust geometry is assumed to be the screen model.

The morphological types of model galaxies were determined by
the bulge-to-total luminosity ratio in the $B$-band.
In this paper we follow the criteria of \cite{Simien1986}:
galaxies with B/T $>$ 0.6, 0.4 $<$ B/T $\leq$ 0.6, and B/T $\leq$ 
0.4 are classified as elliptical, lenticular, and spiral galaxies, 
respectively. 
According to \cite{Kauffmann1993a} and \cite{Baugh1996},
this classification reproduces well the observed type mix.

\section{Parameter settings}
\label{sec:DetParams}
As described in section \ref{sec:model}, 
our model is constructed from physically motivated prescriptions of
several astrophysical processes.
However, a number of free parameter remain.
Here we describe the parameter setting procedure.

\subsection{Overview of parameter settings}
For the cosmological parameters, we adopt the Planck cosmology \citep{Planck2014}.
The several free parameters related to
astrophysical processes are listed in table \ref{tb:params}.
Seven of these parameters, namely,
$\alpha_{\rm star}$, $\tau_{\rm star}$, $\alpha_{\rm hot}$, $V_{\rm hot}$ 
$\alpha_{\rm cool}$, $\varepsilon_{\rm SMBH}$ and $M_{\rm seed}$ were 
tuned to fit the local optical ($r$-band) and near IR ($K$-band) LFs 
and the local mass function (MF) of cold neutral hydrogen, by using a MCMC 
method (see next section).
We use the local LFs and \HI MF as the fiducial references in model calibration,
since they are robustly determined from recent large and deep surveys. 
The other parameters, $f_{\rm bulge}$, $f_{\rm mrg}$, $f_{\rm BH}$, $\tau_{V0}$ 
and $\kappa_{\rm diss}$, are manually tuned by comparing other
observations, since they cannot be constrained by
the local LFs and \HI MF.

The galaxy merger-related parameters, $f_{\rm bulge}$ and 
$f_{\rm mrg}$, are closely related to the abundance of elliptical galaxies,
hence they can be constrained by the LFs divided by morphological class.
However there are still some uncertainties in 
the determination of morphology, thus we did not use them in the fitting.
In this paper we simply assumed that $f_{\rm bulge} = 0.1$ and 
$f_{\rm mrg} = 0.8$, which is the same value with N05.
The mass fraction accreted by SMBH during a starburst, $f_{\rm BH}$, 
affects the bright-end shape of LFs through the AGN feedback; however
it is degenerated with other AGN feedback-related parameters, 
$\varepsilon_{\rm SMBH}$ and $M_{\rm seed}$, and is poorly constrained by LFs.
Thus we tuned $f_{\rm BH}$ to reproduce the observed BH mass -- bulge mass 
relation and mass function of SMBHs, which are significantly affected 
by $f_{\rm BH}$ but not by other two parameters.
We have found that $f_{\rm BH} = 0.005$ is suitable to reproduce the observations
in the case of $f_{\rm bulge} = 0.1$ and $f_{\rm mrg} = 0.8$ 
(see section~\ref{sec:SMBHobs}).
The coefficient of dust extinction, $\tau_{V0}$, was set to the value 
adopted in N05, namely $\tau_{V0} = 2.5 \times 10^{-9}$.
The parameter related to the energy loss fraction in a major merger 
($\kappa_{\rm diss} = 2.0$) was chosen to fit the size--magnitude relation of 
elliptical galaxies (see section \ref{sec:localresults}).
Throughout this paper, we adopt the Chabrier IMF in the mass range 
0.1--100 $M_{\odot}$.

Although the MCMC method is numerically economical, it still requires 
approximately $\sim 10^5$ realizations to estimate the reliable parameter range.
Therefore, to restrain the runtime of each realization within a few seconds, 
we employed the $\nu^2$GC--SS model for the $N$-body data in the MCMC fitting, 
which has the lowest mass resolution and the smallest box size.
The mass resolution of $N$-body data could have complicated effects on the merging 
history of DM halos, thus there is no guarantee that the parameters tuned for $
\nu^2$GC--SS model work well for other $N$-body runs. 
However no significant differences 
between the $\nu^2$GC-SS and $\nu^2$GC--H2 (the highest resolution model) 
are found in the $r$- and $K$-band LFs and \HI MF in the magnitude and mass
range used in the fitting (see section~\ref{sec:fitresult} and 
figure~\ref{fig:LFs}, \ref{fig:HIMF}).

\subsection{Markov chain Monte Carlo analysis}
MCMC analysis was implemented by the Metropolis–-Hastings algorithm 
(\citealt{Metropolis1953}; \citealt{Hastings1970}), 
which is the most commonly used MCMC method.
This method requires the proposal distribution $q$, which
suggests a candidate point for the next step, given the previous sampling point.
We assume a Gaussian distribution function for $q$.
The variance of the Gaussian function is manually selected, 
to decrease the convergence time.
We run 8 MCMC chains in parallel from random starting points.
Each chain has about 50,000 realizations, excluding initial 10,000 steps of
''burn-in'' phase.
The convergence of MCMC chain is checked by the Gelman-Rubin diagnostic test 
(\citealt{Gelman1992}). In this method, the difference between the multiple 
MCMC chains are quantified by the ratio of the variance between chains to the 
variance within chain, $\hat{R}$.
In this paper the chain is considered to have converged when $\hat{R} < 1.1$.
As a result of the fitting, all the free parameters examined here reach convergence.
We simply assume the uniform distribution for the prior probability distribution,
with the range listed in table~\ref{tb:params}.
Although the bounds of prior distributions are physically chosen, 
the ranges are set to be wide in order to cover a large model space,
since our knowledge about the posetrior distribution of parameters is limited.

\subsection{Observational data and error estimation}
\label{subsec:ObsData}
In this subsection, we describe the observational data used in the MCMC fitting.
The local $r$- and $K$-band LFs were obtained by the Galaxy and Mass Assembly (GAMA) 
survey \citep{Driver2012}, and \HI MF was extracted from the data of 
the Arecibo Legacy Fast ALFA (ALFALFA) survey \citep{Martin2010}.

For each realization, the likelihood is calculated as follows:
\begin{equation}
\mathcal{L} = \mathcal{L}_{0}\exp\left
(-\frac{\chi^2_{r}+\chi^2_{K}+\chi^2_{\rm \HI}}{2}
\right),
\end{equation}
where $\mathcal{L}_{0}$ is an arbitrary constant and $\chi_r$, $\chi_K$ and 
$\chi_{\rm \HI}$ are the $\chi^2$ values of the $r$-band LF, $K$-band LF and 
\HI MF, respectively.
These values are estimated as follows:
\begin{equation}
\chi^2 (\phi_{\rm obs}|\theta)= \sum_{i} \frac{\{\phi_{i, {\rm obs}} - \phi_{i, {\rm model}}(\theta)\}^2}{\sigma_{i, {\rm obs}}^2+\sigma_{i, {\rm model}}(\theta)^2}
\end{equation}
where $\phi_{i, {\rm obs}}$ denotes the value in the $i$th bin of the 
observed LF (or \HI MF), $\phi_{i, {\rm model}}(\theta)$ is the value of the model
in the $i$th bin obtained with the parameter set $\theta$ and 
$\sigma_{i, {\rm obs}}$ and $\sigma_{i, {\rm model}}$ are
the errors in the observation and model in each bin, respectively.
The errors in the observed LFs only include Poisson errors (\citealt{Driver2012}), 
while the errors in the observed \HI MF include the systematic errors in mass 
estimation in addition to Poisson errors.

The errors in the model predictions, $\sigma_{i, {\rm model}}$, 
were assumed as the sum of Poisson statistical errors and systematic errors
coming from cosmic variance.
Although the most of SA models assume Poisson errors for the model (e.g., 
\citealt{Henriques2009, Lu2014}), it is controversial whether this assumption 
is appropriate or not.
However typical value of Poisson errors are less than $1\%$ of the error coming
from the cosmic variance described below, it does not have significant effect 
on the parameter fitting.
Although the errors of the observed \HI MF include
the systematic errors come mainly from uncertainty 
of mass estimation, especially for low \HI mass galaxies (see, e.g.,
\citealt{Zwaan2005,Martin2010}), we does not include it 
in the errors of our model.

The effect of cosmic variance is estimated as follows.
First, we ran the model using the $\nu^2$GC-S for the {\it N}-body data,
which has a larger box ($L = 280 h^{-1}$ Mpc) than that in 
the MCMC fitting ($L = 70 h^{-1}$ Mpc), 
and randomly picked out the $L = 70 h^{-1}$ Mpc box 
from the large box data in $\sim$100 trials.
Following this, we drew the LFs and \HI MF from the
small boxes and determined their uncertainties (approximately 20\%) 
in each bin. We accounted for this 20\% uncertainty in $\sigma_{i, {\rm moodel}}$, 
in addition to Poisson errors.

Populations of dwarf galaxies with low surface brightness are known to exist,
and the faint-end slope of observed LFs may be affected from the surface brightness 
limits of galaxy surveys (e.g. \citealt{Blanton2005}).
According to \cite{Baldry2012}, the incompleteness of GAMA samples become 
larger than 30\% at $\mu_{r,50} \gtrsim$ 23.5 mag ${\rm arcsec}^{-2}$, 
where $\mu_{r,50}$ is the surface brightness within the Petrosian half-light radius.
Therefore, we adopt this limit in calculating the model LFs.
To calculate the Petrosian surface brightness, we require the light profile of galaxies.
However, because our model does not resolve the internal structure of galaxies,
we converted the effective radius and total magnitude into the Petrosian radius 
and Petrosian magnitude, respectively, fixing the S\'ersic index $n_{\rm s}$ 
of bulge ($n_{\rm s} = 4$) and disk ($n_{\rm s} = 1$) components for all galaxies.

The mass of cold atomic hydrogen of model galaxies are estimated as follows.
First we assume that the 75\% of the cold gas is composed of hydrogen.
This cold hydrogen will be split into atomic and molecular;
however, our current model does not follow the complex history of 
the formation of molecular hydrogen.
Therefore we simply assume a fixed $\mathrm{H}_{2}$-to-\HI
ratio for all galaxies.
According to the observational estimation of \cite{Keres2003} 
and \cite{Zwaan2005}, 
a global mass ratio of molecular to atomic hydrogen is $\sim0.4$.
Thus the mass of cold atomic hydrogen is estimated as
\begin{equation}
M_{\rm \HI} = 0.75/(1+0.4) M_{\rm cold}.
\end{equation}
Note that similar approach was used in other SA models
(e.g. \citealt{Power2010}; \citealt{Lu2012}).
When fitting \HI MF, we only used the data points acquired at 
$M_{\rm \HI} > 10^{8} M_{\odot}$, because at masses below 
this limit, the mass resolution the {\it N}-body data
would affect the shape of low mass end of \HI MF 
(see next subsection).
The uncertainties in the observed \HI MF also increase
below this limit due to the incompleteness of the survey 
(\citealt{Martin2010}).

\subsection{Fitting results}
\label{sec:fitresult}
\begin{figure*}
    \includegraphics[width=2\columnwidth]{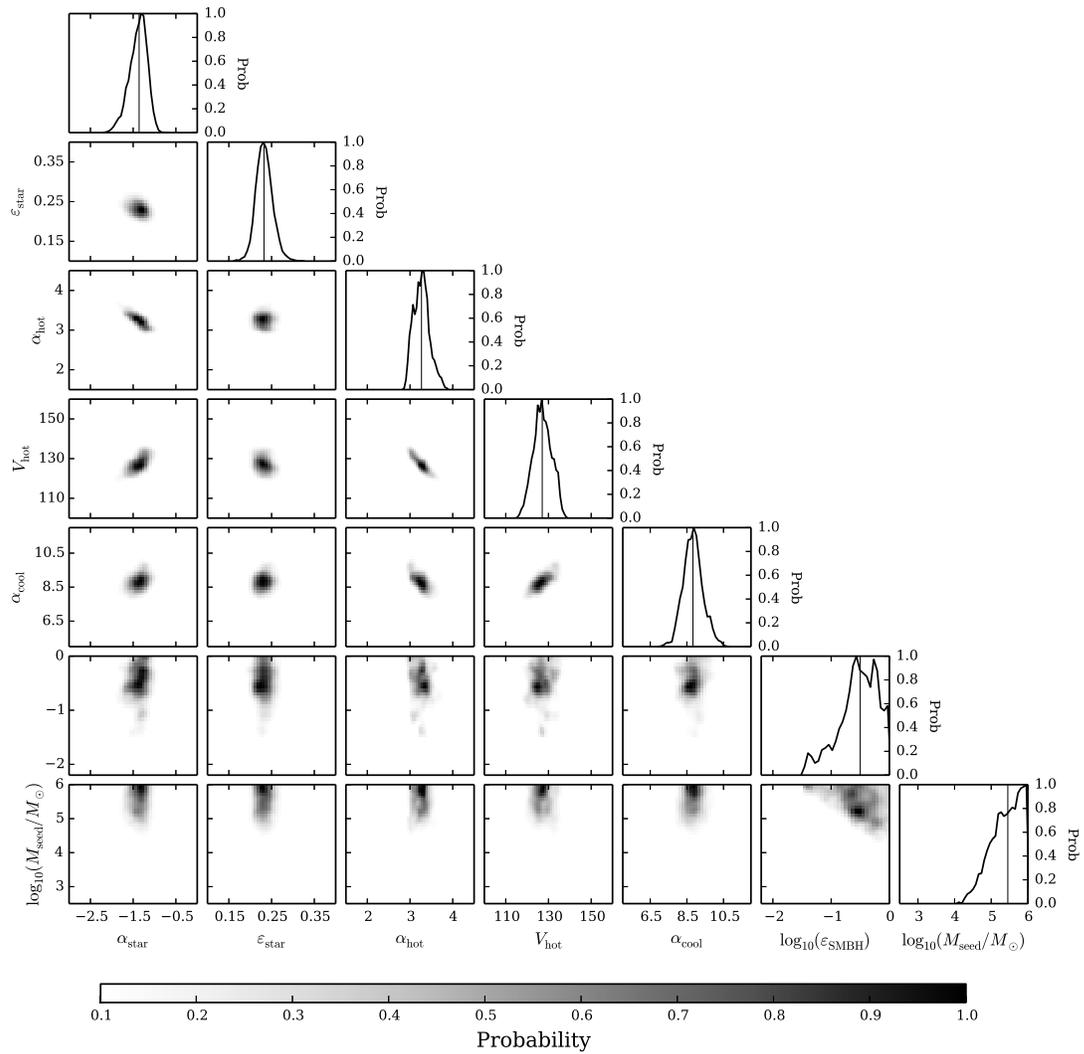}
    \caption{
    The 1D (diagonal panels) and 2D (off-diagonal panels) 
    posterior probability distribution functions
    of five free parameters tuned in MCMC fitting.
    The solid vertical lines in diagonal panels show the median of each distribution.
    Probability distributions of all combinations of the five parameters are shown
    by grey contours.
    }
    \label{fig:params}
\end{figure*}

The diagonal panels of figure~\ref{fig:params} present the 1D posterior 
probability distributions of the parameters tuned in the MCMC fitting. 
From the 1D posterior probability distributions, we computed 
the medians and 10 and 90 percentiles of each parameter; the statistics are 
summarized in table~\ref{tb:params}.
The off-diagonal panels of figure~\ref{fig:params} present the 2D posterior 
probability distributions of all combinations of the seven free model parameters (grey contours).
The 1D distributions of the five parameters, $\alpha_{\rm star}$, 
$\tau_{\rm star}$, $\alpha_{\rm hot}$, $V_{\rm hot}$, and $\alpha_{\rm cool}$ are
highly-peaked, indicating that they are well constrained within the assumed range.
On the other hand, $\varepsilon_{\rm SMBH}$ and $M_{\rm seed}$ have broad 
distribution. This can be understood as follows.
If these parameters are large enough, the second condition of 
AGN feedback [equation~(\ref{eq:AGNe})] will be satisfied in all halos. 
In such case, the specific value of these parameters no longer affect the shape of 
LFs and MF, and therefore only the lower boundary of these can be constrained 
from the fitting of LFs and \HI MF.
The posterior distribution of $M_{\rm seed}$ suggests that the $M_{\rm seed}$ 
should be larger than $10^{5} M_{\odot}$. This seed BH mass is somewhat higher
than other SA models.
In our model a fraction of central galaxies is bulge-less (i.e. they have never 
experience any major merger event till $z = 0$). The SMBH mass is equal to $M_{\rm 
seed}$ in such galaxies, thus large value of $M_{\rm seed}$ is required in order to 
make AGN feedback work in such halos.

\begin{figure*}
    \subfigure{\includegraphics[width=1.0\columnwidth]{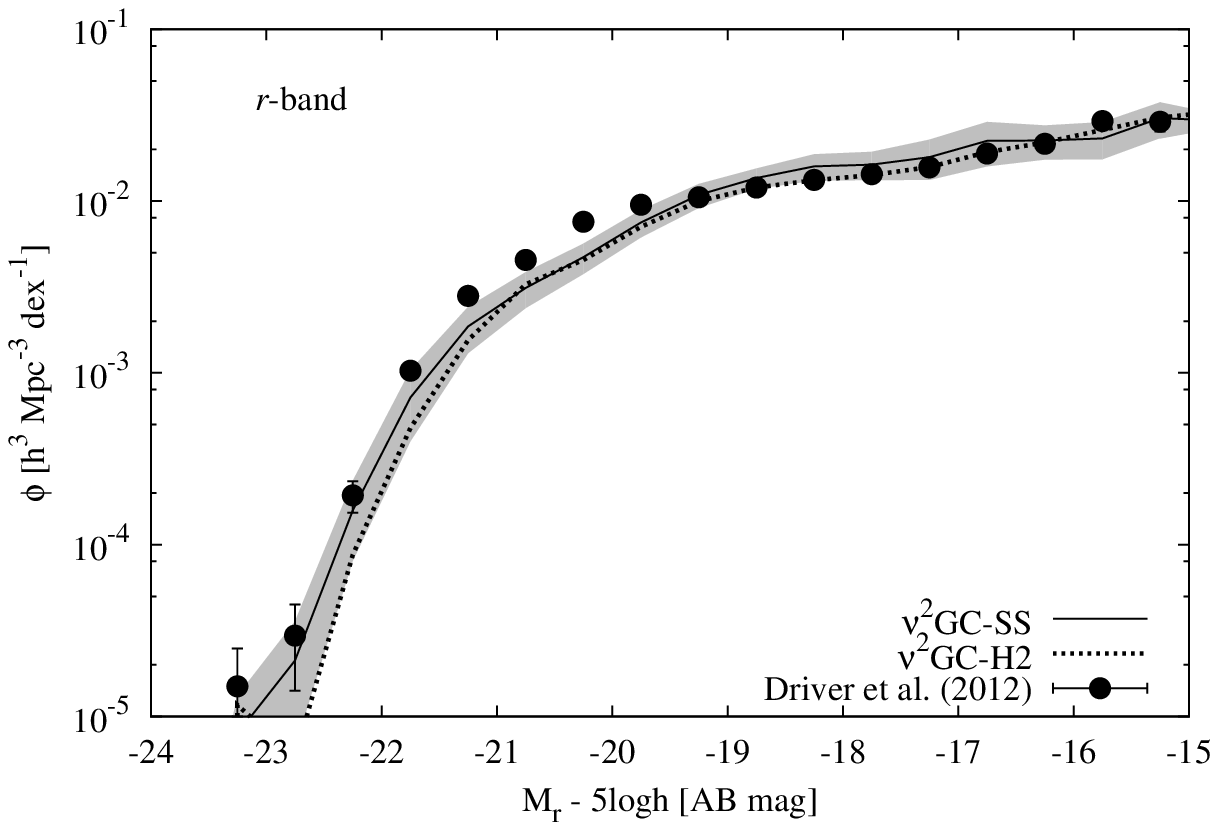}}
    \subfigure{\includegraphics[width=1.0\columnwidth]{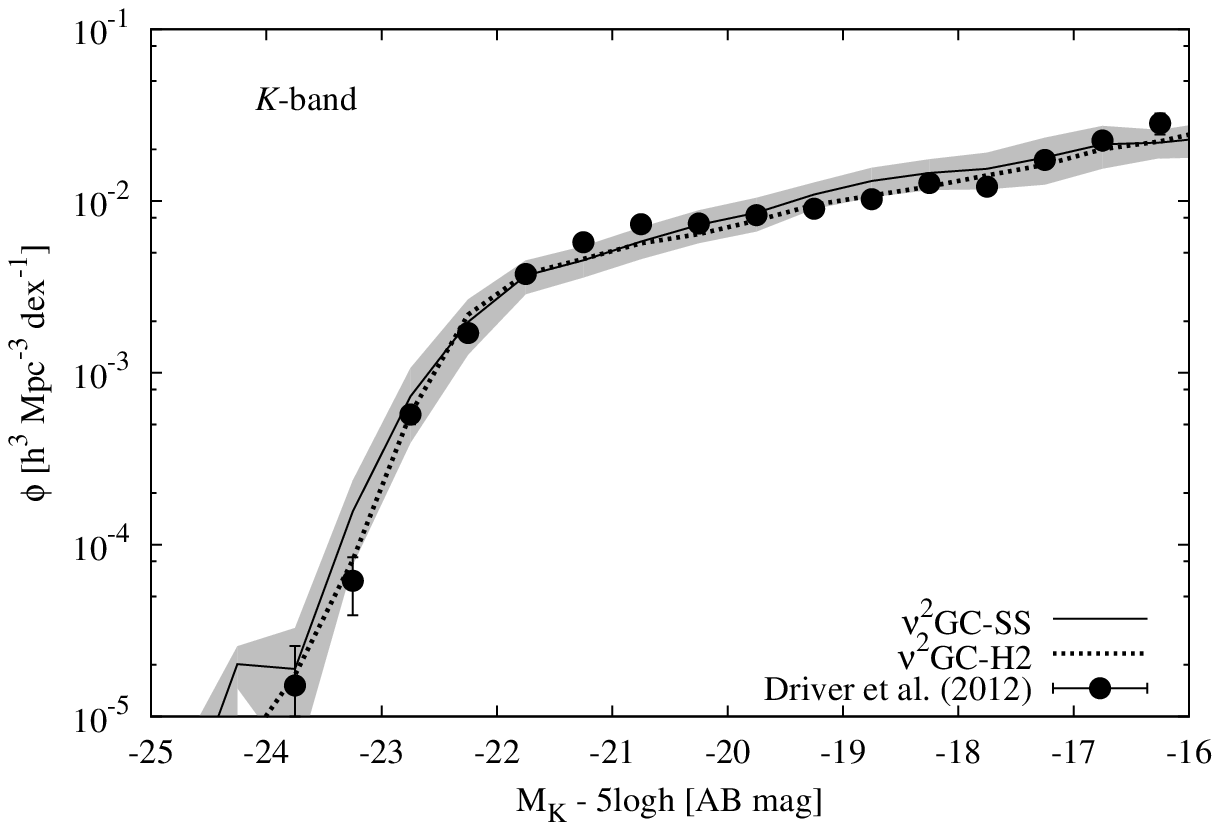}}
    \caption{The $r$- and $K$-band LFs. 
    The black solid line represents the model ($\nu^2$GC-SS) with the best-fit 
    parameters determined by MCMC fitting.
    The black dotted line shows the model with the same parameters but using 
    the high-resolution {\it N}-body data ($\nu^2$GC-H2).
    The shaded region denotes the $1\sigma$ error in the model, estimated from the posterior 
    probability distribution of the parameters (figure \ref{fig:params}).
    The observational data shown in black filled circles are obtained by GAMA survey 
    (Driver et al. 2012).}
    \label{fig:LFs}
\end{figure*}

Figure \ref{fig:LFs} present the $r$- and $K$-band LFs in the model with 
the MCMC-obtained best fit parameters. 
The model closely matches the observations over all magnitude ranges.
The shaded regions indicate the $1 \sigma$ error 
in the model LFs, estimated from the $1 \sigma$ confidence interval of each parameter.
To see the effect of the mass resolution, we also plot the results of 
$\nu^2$GC-H2 model with the same parameters.
These models are consistent within the $1 \sigma$ error.

\begin{figure}
    \includegraphics[width=\columnwidth]{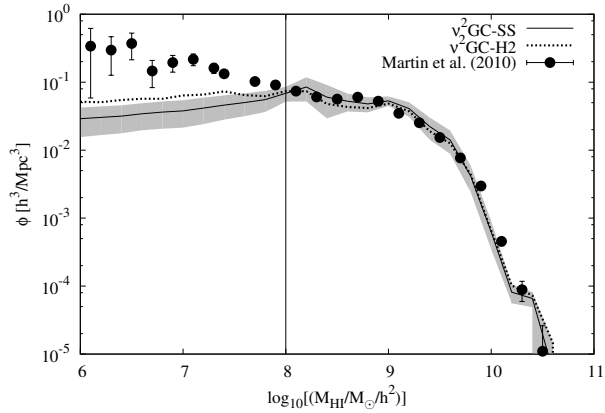}
    \caption{\HI mass function of the best fit model.
    The black solid line represents the model ($\nu^2$GC-SS) with the best-fit 
    parameters determined by MCMC fitting.
    The black dotted line shows the model with the same parameters 
    but using the high-resolution {\it N}-body data ($\nu^2$GC-H2).
    The mass resolution of {\it N}-body data would affects the shape of MF
    below $M_{\rm \HI} \sim 10^{8} M_{\odot}$ (shown by vertical solid line);
    therefore, data below this were excluded in the parameter fitting.
    The shaded region denotes the $1\sigma$ error in the model, estimated from the posterior 
    probability distribution of the parameters (figure \ref{fig:params}).
    Black filled circles are the observational data obtained by ALFALFA survey
    (\citealt{Martin2010}).
    }
    \label{fig:HIMF}
\end{figure}

Figure \ref{fig:HIMF} shows \HI MF computed by the best-fit model.
Data below the lower limit of the \HI mass (solid vertical line) were 
excluded in the MCMC fitting because they deviated when the model was run at 
higher mass resolution ($\nu^2$GC-H2 model, $M_{\rm min} = 1.37\times10^8 
h^{-1}M_{\odot}$; dashed line).
Although there remains uncertainties in both the model and the observation,
the model seems to underpredict the abundance of lower \HI mass
galaxies ($M_{\rm \HI} < 10^{8} \Msun$).
The similar trend is seen in other SA models.
For example, \cite{Gonzalez-Perez2014} find that their model also 
underpredicts the abundance of galaxies at the lower-mass end of
\HI MF (see also \citealt{Lagos2014}).
They conclude that this is mainly due to the limited mass resolution
of their {\it N}-body data.
However, even in the $\nu^2$GC-H2 model, which has approximately
two orders of magnitude higher mass resolution than that of
\cite{Gonzalez-Perez2014}, the lower-mass end of the \HI MF is 
still underpredicted.
This result might suggest that the more realistic modeling of
star formation and SN feedback is required (e.g., \citealt{Lu2014}; \citealt{Benson2014}).
Furthermore,  non-virialized gas which is not included in 
our model and/or \HI gas with low \HI column densities below 
the observation limits of the current \HI blind surveys might contribute to the low end of the \
HI MF (e.g., \citealt{Okoshi2010}).
We will further investigate this issue in the future.

\begin{table*}[]
\caption{The list of free parameters related to astrophysical processes.
Seven of these parameters, namely,
$\alpha_{\rm star}$, $\tau_{\rm star}$, $\alpha_{\rm hot}$, $V_{\rm hot}$,
$\alpha_{\rm cool}$, $\varepsilon_{\rm SMBH}$, and $M_{\rm seed}$were 
tuned to fit the local LFs and \HI MF, by using a MCMC method.
See text for details of parameter settings.
}
\centering
  \begin{tabular}{lcccl}
  \hline
   &       &\multicolumn{2}{c}{Posterior}    & \\
   & Prior & Median & 10 to 90 percentile & Meaning\\
   \hline
  $\alpha_{\rm star}$        & [-5.0, 0.0]   & -1.36 & [-1.14, -1.67] & star formation-related\\
  $\varepsilon_{\rm star}$   & [0.01, 1.0]   & 0.23 & [0.21, 0.26] & star formation-related\\
  $\alpha_{\rm hot}$         & [0.0, 5.0]    & 3.27 & [3.03, 3.52] & SN feedback-related \\
  $V_{\rm hot}$ (km/s)       & [50.0, 200.0] & 127.1 & [121.6, 133.1] & SN feedback-related\\
  $\alpha_{\rm cool}$        & [0.1, 10.0]   & 8.83 & [8.16, 9.55] & AGN feedback-related\\
  $\log_{10}(\varepsilon_{\rm SMBH})$ & [-2.0, 0.0] & -0.50 & [-1.02, -0.14] & AGN feedback-related\\
  $\log_{10}(M_{\rm seed}/M_{\odot})$ & [3.0, 6.0] & 5.45 & [4.83, 5.90] &seed black hole mass \\
  $f_{\rm BH}$               & -- & $5\times10^{-3}$ (fix) & -- & fraction of the mass accreted onto SMBH \\
  &&&&during major merger\\
  $\tau_{V0}$     & -- & $2.5\times10^{-9}$ (fix) & -- & coefficient of dust extinction\\
  $f_{\rm bulge}$ & -- & 0.1 (fix) & -- & major/minor merger criterion\\
  $f_{\rm mrg}$   & -- & 1.0 (fix) & -- & coefficient of dynamical friction time scale\\
  $\kappa_{\rm diss}$ & -- & 2.0 (fix) & -- & energy loss fraction\\
  \hline 
  \end{tabular}
  \label{tb:params}
\end{table*}
\section{Numerical galaxy catalog}
\label{sec:catalog}
Following the above procedures, we finally obtained
the numerical galaxy catalog.
This catalog contains various data on each mock galaxies:
redshift, three-dimensional positions, physical quantities such as stellar mass, 
gas mass, metallicity, star formation rate, effective radius, 
and magnitudes in several passbands in the UV--NIR regime.
More information is provided at here\footnote[1]{}.

\begin{figure*}
    \begin{center}
    \includegraphics[width=2\columnwidth]{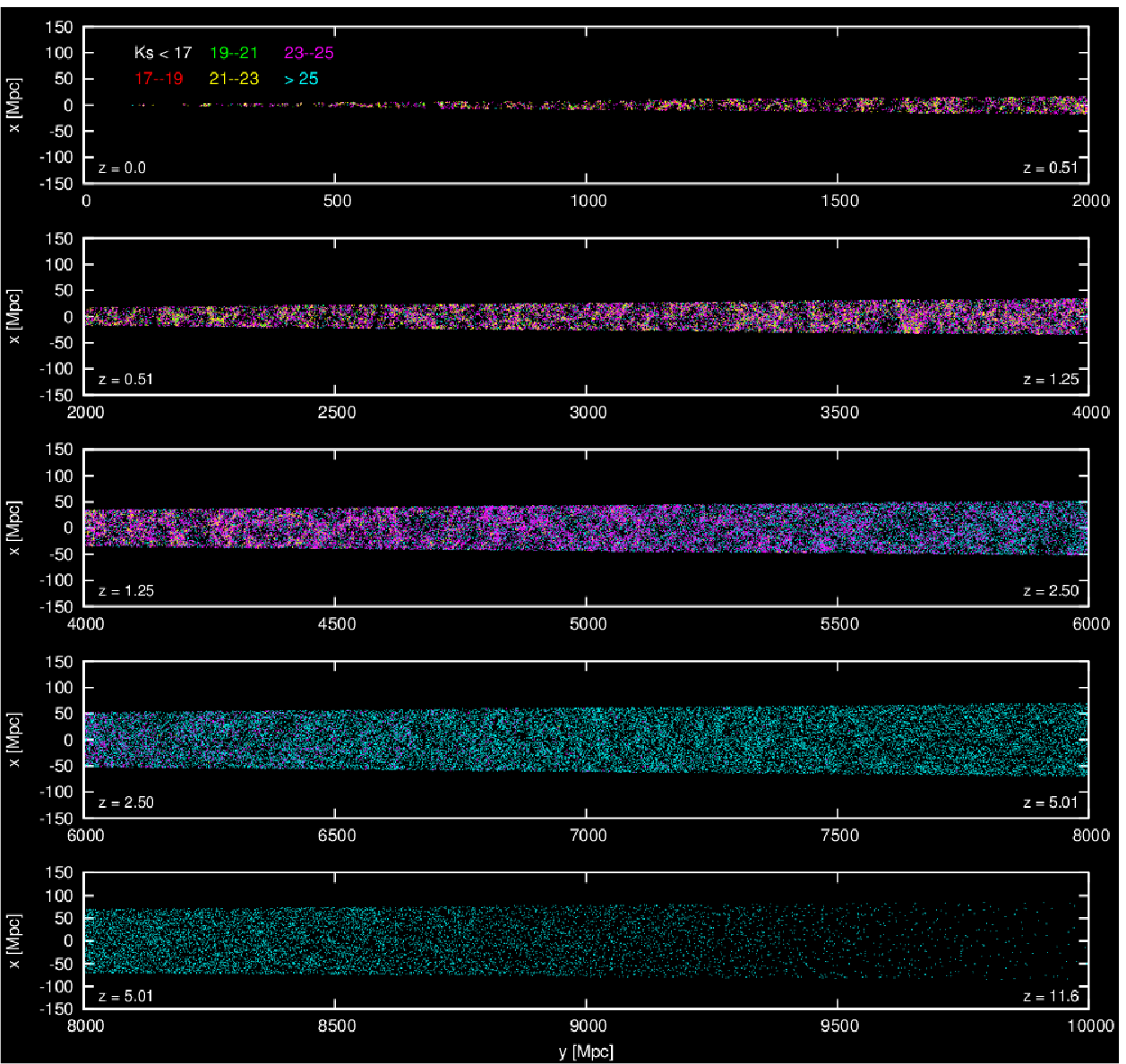}
    \end{center}
    \caption{
    The spatial distribution of the mock galaxies plotted on the past light-cone 
    of an observer located at redshift zero. 
    The color indicates the apparent magnitude of each galaxies in the 2MASS Ks-band.
    We only show the one-thousandth of galaxies which are randomly picked up 
    from total sample, to avoid a confusion. 
    }
    \label{fig:map}
\end{figure*}

Figure \ref{fig:map} plots the spatial distribution of the model galaxies
from $z = 0.0$ to $z = 11.6$ (corresponding to approximately $10^4$~Mpc along 
the comoving radial distance), plotted on the past light-cone of an observer at 
$z = 0$.
Here we show the result of the $\nu^2$GC-H1 model.
The light-cone is generated by patchworking the model outputs at 
various redshift slices. During the patchworking,
the simulation box was randomly shifted and rotated to avoid artifacts in 
the spatial structure. 
Web-like structures are clearly visible in this figure. 
Thanks to the high mass resolution of the model, we can observe 
galaxies even at $z > 10$.

\section{Local galaxies}
\label{sec:localresults}
In this section we compare the model predictions with local observations.
In what follows, we show the results of the $\nu^2$GC-H2 model 
which has the highest mass resolution, 
unless otherwise mentioned.
The adopted parameters related to the baryon physics are listed in table \ref{tb:params}.

\subsection{Size and disk rotation velocity of spiral galaxies}
First, we compare the predicted effective radius and disk rotation velocity
of spiral galaxies with the observations.
For the observational data, we use the data taken from \cite{Courteau2007}.
The sample of \cite{Courteau2007} is a compilation of the 
major samples of local spiral galaxies for which rotational velocities
are available. Their sample includes \cite{Mathewson1992}, 
\cite{Dale1999}, \cite{Courteau2000}, \cite{Tully1996} and \cite{Verheijen2001}.
The disk scale length of the sample galaxies are estimated from
the $I$-band image, and the disk rotation velocities are estimated from 
H$\alpha$ or \HI line widths.
Both of the disk size and the rotation velocity are corrected for inclination.

\begin{figure}
    \includegraphics[width=\columnwidth]{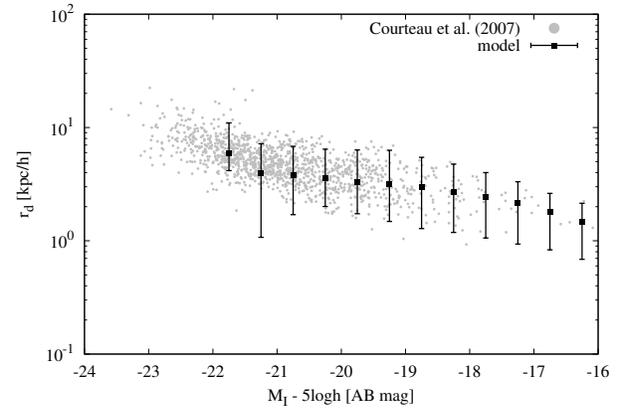}
    \caption{Effective radius of spiral galaxies plotted against $I$-band magnitude.
    The black filled squares with error bars show the median and the 10th to 
    the 90th percentile of 
    the predicted size of model galaxies in each magnitude bin.
    Small gray dots are the observational data obtained by Courteau et al. (2007).}
    \label{fig:SM_S}
\end{figure}

\begin{figure}
    \includegraphics[width=\columnwidth]{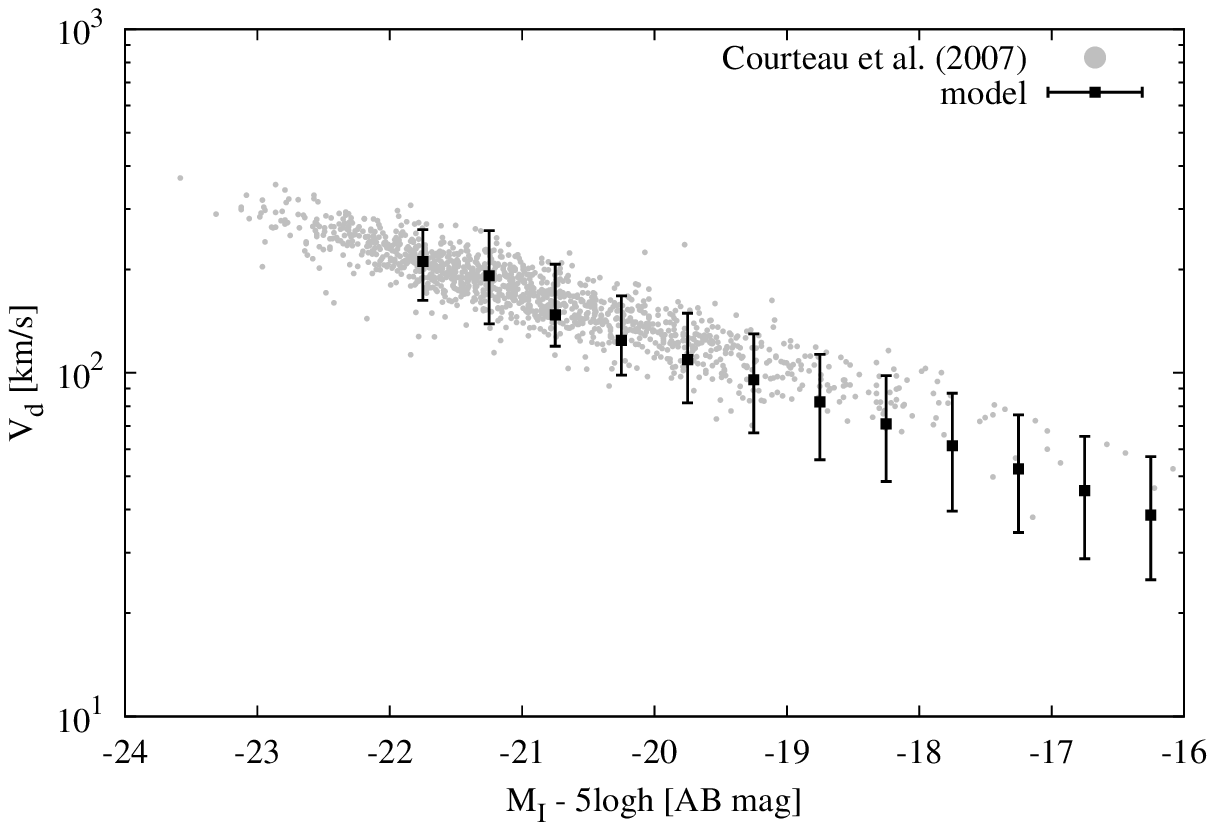}
    \caption{$I$-band Tully-Fisher relation (i.e., disk rotation velocity against
    $I$-band magnitude) of spiral galaxies. 
    The black filled squares with error bars show the median and the 10th to 
    the 90th percentile of 
    the predicted disk velocity of model galaxies in each magnitude bin.
    Small gray dots are the observational data obtained by Courteau et al. (2007).
    }
    \label{fig:TF}
\end{figure}

Figures \ref{fig:SM_S} and \ref{fig:TF} show the scaling relations 
between the effective disk radius and the $I$-band magnitude, and
between the disk rotation velocity and the $I$-band magnitude 
(the so-called Tully-Fisher relation; \citealt{Tully1977}),
respectively.
The median and the 10th to the 90th percentile of the distribution of
model galaxies in each magnitude bin are shown by black squares with error bars.
The observational data are shown by small gray dots.
The model very well reproduces these observed scaling relations 
over all magnitude range.
The effect of the dynamical response to the disk will be investigated
in detail in future paper.

\subsection{Size and velocity dispersion of elliptical galaxies}
\label{sec:Esize}
In this subsection we compare the predicted effective radius and velocity dispersion
of elliptical galaxies with the observations.
For the observational data, we use the data compiled by \cite{Forbes2008}.
They take the central velocity dispersions of sample galaxies from
the catalog of \cite{Bender1990}, \cite{Bender1992}, 
\cite{Burstein1997}, \cite{Faber1989}, \cite{Trager2000},
\cite{Moore2002}, \cite{Matkovic2005} and \cite{Firth2007}.
The half-light radii are calculated from 2MASS $K$-band 20th isophotal size,
by using a empirical relation based on S\'ersic light profiles
(\citealt{Forbes2008}).

\begin{figure}
    \includegraphics[width=\columnwidth]{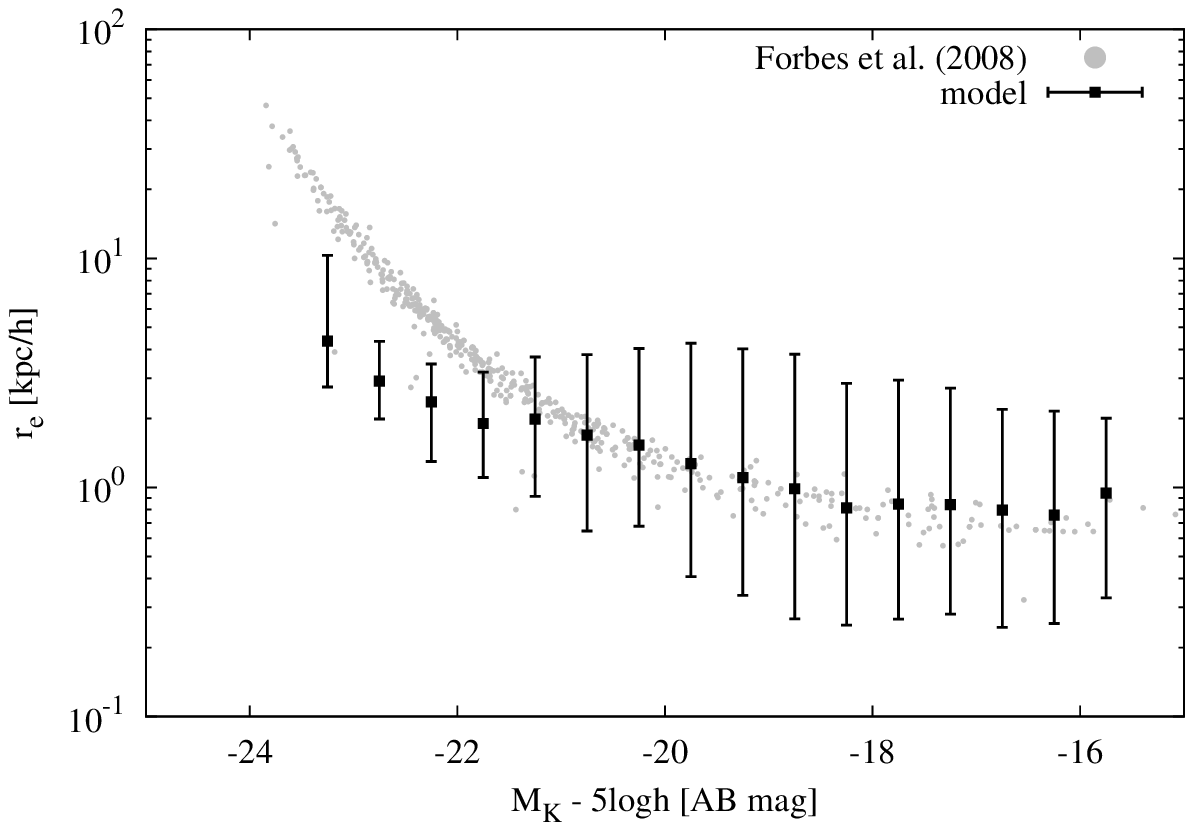}
    \caption{Effective radius of elliptical and lenticular 
    galaxies plotted against $K$-band magnitude.
    The black filled squares with error bars show the median and the 10th 
    to the 90th percentile of 
    the predicted size of model galaxies in each magnitude bin.
    Small gray dots are the observational data estimated by Forbes et al. (2008).
    }
    \label{fig:SM_E}
\end{figure}

\begin{figure}
    \includegraphics[width=\columnwidth]{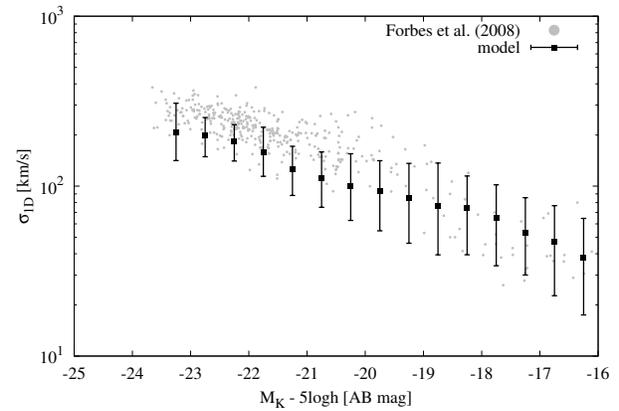}
    \caption{$K$-band Faber-Jackson relation (i.e., projected central velocity 
    dispersion against $K$-band magnitude) of elliptical and lenticular galaxies.
    The black filled squares with error bars show the median and the 10th to 
    the 90th percentile of 
    the distribution of predicted velocity dispersion of model galaxies 
    in each magnitude bin.
    The projected velocity dispersion of model galaxies are estimated as 
    $\sigma_{\rm 1D} = V_{\rm b}/\sqrt{3}$ after being increased to the central
    value by a factor of $\sqrt{2}$ assuming the de Vaucouleurs profile.
    Small gray dots are the observational data compiled by Forbes et al. (2008).
    }
    \label{fig:FJ}
\end{figure}

Figures \ref{fig:SM_E} and \ref{fig:FJ} show the scaling relations 
between the effective radius and the $K$-band magnitude, and
between the velocity dispersion and the $K$-band magnitude 
(the so-called Faber-Jackson relation; \citealt{Faber1976}),
respectively.
The median and the 10th to the 90th percentile of the distribution of
model galaxies in each magnitude bin are shown by black squares with error bars.
The effective radius of model galaxy, $r_{e}$, is estimated 
from $r_{e}=0.744 r_{b}$ \citep{Nagashima2003}, where $r_{b}$ is 
the three-dimensional half-mass radius.
The projected velocity dispersion is estimated as $\sigma_{\rm 1D} 
= V_{\rm b}/\sqrt{3}$ after being increased to the central value 
by a factor of $\sqrt{2}$ assuming the de Vaucouleurs profile.

As shown in figures \ref{fig:SM_E} and \ref{fig:FJ},
our model underpredicts both of the size and the velocity dispersion 
of galaxies brighter than $M_{\rm K} \sim -20$, comparing with the observations. 
The size and velocity dispersion are related to the dynamical mass of a galaxy
as $M_{\rm dyn} \propto r_{\rm e}^{2} V_{\rm b}$,
and therefore the model also underpredicts the dynamical mass of 
bright elliptical galaxies at a fixed magnitude.
These results might imply that our treatment of bulge (and elliptical galaxies) 
formation process is oversimplified.
We need to consider more realistic model for galaxy merger, as well as 
another channels of bulge formation such as disk instabilities.
Furthermore, assumed IMF might also be responsible for the underprediction
of mass-to-luminosity ratio.

\subsection{Cold gas}
\begin{figure}
    \includegraphics[width=\columnwidth]{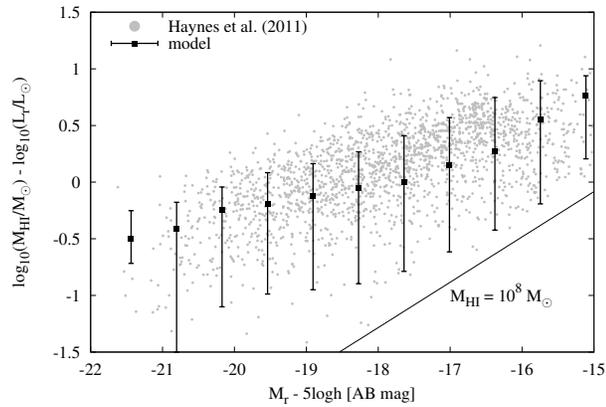}
    \caption{Cold gas mass relative to $r$-band luminosity as a 
    function of $r$-band magnitude for spiral galaxies.
    Small gray dots are the observational data obtained 
    by 40\% catalog of ALFALFA survey ($\alpha.40$; Haynes et al. 2011). 
    Here we only show the galaxies having \HI mass greater than 
    $10^8 \Msun$.
    The solid diagonal line corresponds to the constant 
    hydrogen mass of $M_{\rm \HI} = 10^8 M_{\odot}$.
    The black filled squares with error bars show the median and the 10th to the 
    90th percentile of distributions of the model galaxies in each magnitude bin.
    We simply estimated the mass of cold atomic hydrogen as 
    $M_{\rm \HI} = 0.54 M_{\rm cold}$ (see text for detail).
    }
    \label{fig:MrMHI}
\end{figure}

Figure \ref{fig:MrMHI} presents the cold atomic hydrogen mass relative 
to the $r$-band luminosity against the $r$-band magnitude for local 
spiral galaxies.
As described above, the atomic hydrogen mass of model galaxy is estimated as 
$M_{\rm \HI} = 0.54M_{\rm cold}$ (see section \ref{subsec:ObsData}).
The median and the 10th to the 90th percentile of the distribution of
model galaxies in each magnitude bin are shown by black squares with error bars.
The observational data shown in small gray dots are taken from 
ALFALFA 40\% catalog ($\alpha.40$; \citealt{Haynes2011}).

As already mentioned above, the uncertainties in the model 
increase for galaxies having \HI mass less than $10^8 \Msun$
(see section \ref{subsec:ObsData}).
Furthermore, the $\alpha.40$ catalog is highly incomplete
for galaxies at $M_{\rm \HI} < 10^8 \Msun$ (\citealt{Haynes2011}).
Therefore we only plot galaxies having $M_{\rm \HI} > 10^8 \Msun$
for both of the model and the observation.
The diagonal solid line in figure \ref{fig:MrMHI} corresponds
to $M_{\rm \HI} = 10^8 \Msun$.

We can see that the model very well reproduces the observation
over all magnitude range.
The cold gas mass to luminosity ration is mainly determined 
by the balance of gas consumption rate by star formation and SN feedback.
The agreement between the model and observation seen in 
figure \ref{fig:MrMHI} supports the validity of our model of
star formation and SN feedback.
For more detailed discussion on the properties of cold gas component in
our model, we refer the reader to \cite{Okoshi2005} and \cite{Okoshi2010}
although they are based on our previous model.


\subsection{Supermassive black holes}
\label{sec:SMBHobs}
\begin{figure}
    \includegraphics[width=\columnwidth]{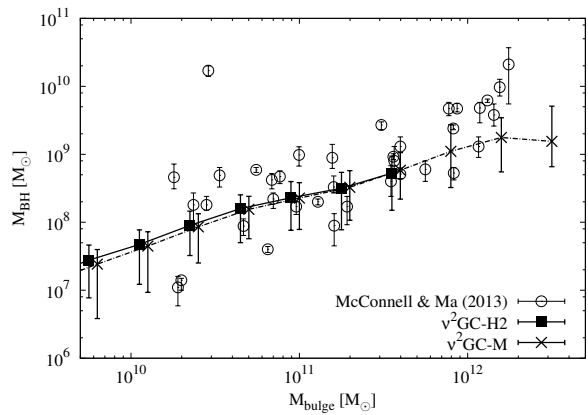}
    \caption{The SMBH mass versus bulge mass relation.
        The black filled squares with error bars show the median and the 10th to 
        the 90 percentile
        of the distribution of model galaxies in each bin of bulge mass for 
        the $\nu^{2}$GC-H2 model. The crosses are the results of $\nu^2$GC-M
        model, which is shifted about $+0.05$ dex in $\log(M_{\rm bulge}/\Msun)$ to avoid
        a confusion.
        The observational data obtained by \cite{McConnell13} are shown by
        open circles.}
    \label{fig:mbhmbulge}
\end{figure}

\begin{figure}
    \includegraphics[width=\columnwidth]{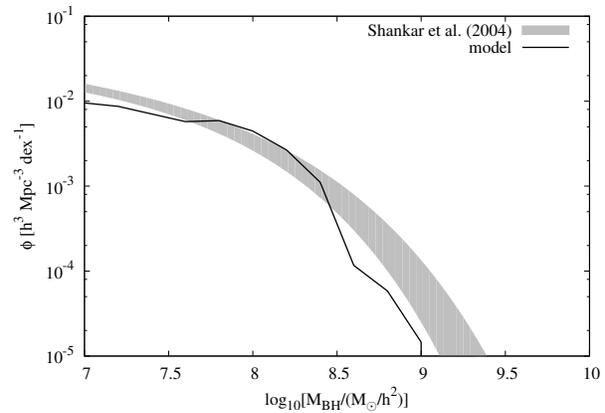}
    \caption{The mass function of local SMBHs.
        The analytical fit to the observational data obtained by 
        \cite{Shankar2004} is shown by the gray shaded region.
        The black solid line is the best-fit model.}
    \label{fig:smbhmf}
\end{figure}

In this subsection we present the properties of SMBHs at local universe.
Figure \ref{fig:mbhmbulge} shows the predicted relation between
the bulge mass and the SMBH mass, comparing with the observational 
data obtained by \cite{McConnell13}.
To show the distribution of more massive and rarer objects,
we also plot the $\nu^2$GC-M model in this figure.
With a fixed mass fraction of cold gas accreted by a SMBH during a starburst
($f_{\rm BH} = 0.005$), the observed relation is naturally explained by the model.
However $f_{\rm BH}$ degenerates with other parameters which are related to
bulge formation and SMBH evolution, such as $f_{\rm bulge}$ and $f_{\rm mrg}$, 
and therefore we need another observational constraints to discuss the physical 
meanings of these parameters.
For example, morphology-dependent LFs will help to resolve the degeneracy
since $f_{\rm bulge}$ and $f_{\rm mrg}$ control the abundance of bulge component.
Gravitational waves from SMBHs will also provide us strong and 
independent constraints (see, e.g., \citealt{Enoki2004};
\citealt{Enoki2007}).
Figure \ref{fig:smbhmf} show the predicted MF of local SMBHs,
comparing with the observational estimation by \cite{Shankar2004}.
The model also well reproduces the observation over all mass range.

For more detailed discussions on the properties of AGN populations, 
see \cite{Enoki2003}, \cite{Enoki2014} and \cite{Shirakata2015}, 
although they are based on our previous model.

\subsection{Distributions of galaxy colors}
\begin{figure}
    \begin{center}
    \includegraphics[width=\columnwidth]{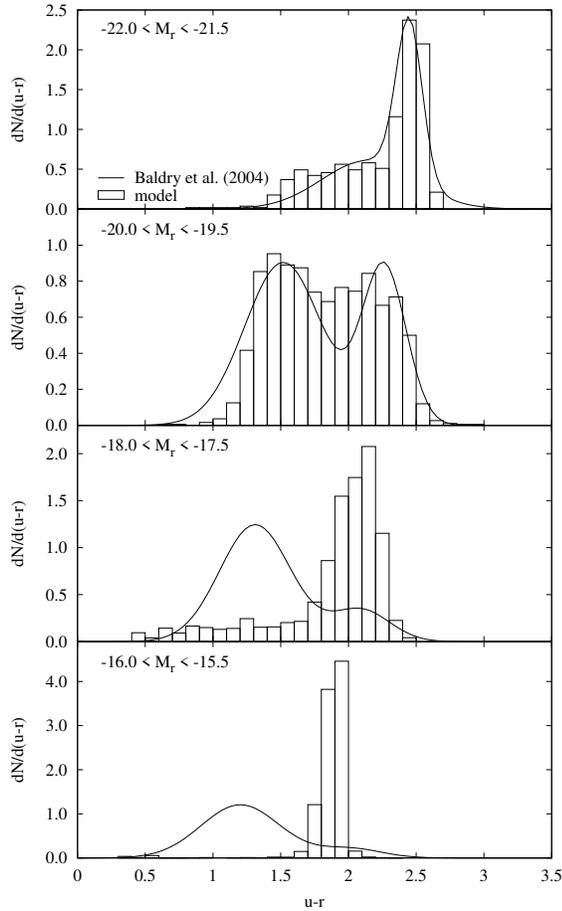}
    \end{center}
    \caption{Color distribution of galaxies (i.e., differential 
    number of galaxies per color bin) in the different $r$-band 
    magnitude bins (from top to bottom, $-22.0 < M_{\rm r} < -21.5$, 
    $-20.0 < M_{\rm r} < -19.5$, $-18.0 < M_{\rm r} < -17.5$ and
    $-16.0 < M_{\rm r} < -15.5$). 
    The black solid lines in each panel are the analytical
    fit to the distribution of SDSS galaxies obtained by \cite{Baldry2004}.
    The black histograms are the model predictions.
    Both the model and observation are normalized to unity.}
    \label{fig:ColorDist}
\end{figure}

Figure \ref{fig:ColorDist} shows the distributions of 
$(u-r)$ color of galaxies (i.e., differential number of galaxies per color bin) 
divided in several bins of the $r$-band magnitude.
We compare the model predictions with the observed distributions 
extracted from Sloan Digital Sky Survey (SDSS) catalog \citep{Baldry2004}.
As shown in figure \ref{fig:ColorDist}, the model well reproduces 
the observed bimodal distributions for the galaxies brighter than
$M_{\rm r} = -19.5$.
However the model predicts systematically redder color for faint galaxies.
This result might imply that the faint galaxies in our model obtain its stellar mass
too early and have exhausted the almost all of cold gas, and consequently have
redder colors.
This discrepancy would be due to the oversimplified modeling of 
the star formation, SN feedback and stripping of hot gas in subhalos
(cf. \citealt{Makiya2014}).
We will investigate this issue in future paper. 

\subsection{Main sequence of star-forming galaxies}
It has been known that the SFR and the stellar mass of star-forming 
galaxies are tightly correlated (the so-called ``star-forming main sequence'';
e.g., \citealt{Brinchmann2004}; \citealt{Elbaz2007}; \citealt{Salim2007};
 \citealt{Daddi2007}).
 
\begin{figure}
    \includegraphics[width=\columnwidth]{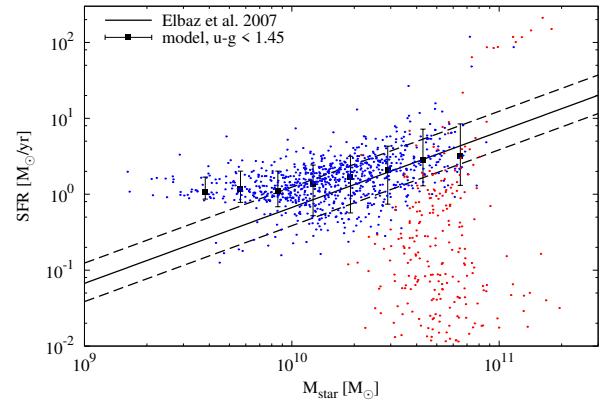}
    \caption{The stellar mass vs SFR relation for local galaxies. 
    Both the SFR and stellar mass are converted into those with Salpeter IMF
    from those with Chabrier IMF, by multiplying by a factor of 1.8.
    The solid and dashed lines are the observed relation and typical error
    obtained by \cite{Elbaz2007}.
    The blue and red dots show the distribution of star-forming 
    and passive galaxies in the model, respectively.
    Here we adopt the same color criteria with \cite{Elbaz2007},
    i.e., the galaxies having blue color ($(u-g) < 1.45$) are regarded 
    as the star-forming while the others are regarded as passive.
    For the model, we only plot the galaxies brighter than $M_{B} = -20$ AB mag,
    which is the limiting magnitude of the sample of \cite{Elbaz2007}.
    The black filled squares with error bars show the median and the 10th to 
    the 90th percentile of star-forming galaxies in each bin of stellar mass.
    }
    \label{fig:MainSequence}
\end{figure}

Figure \ref{fig:MainSequence} shows the SFR against 
the stellar mass for the model galaxies at $z = 0$.
The star-forming galaxies and passive galaxies are shown in
blue and red dots, respectively.
The black squares with error bars show the median and the 10th to the 90th 
percentile of the distributions of star-forming galaxies in each stellar mass bin.
In the same figure we also show the observed relation obtained by
\cite{Elbaz2007} by a solid line with typical errors by dashed lines.
For the model galaxies, we adopt the same
limiting magnitude, $M_{B} < -20$ AB mag, with the sample
of \cite{Elbaz2007}. The separation criteria between the star-forming
galaxies and passive galaxies is also the same with \cite{Elbaz2007}:
the galaxies having $(u-g) < 1.45$ are star-forming and the others are
passive.
Both of the SFR and the stellar mass are converted into those with Salpeter 
IMF from those with Chabrier IMF, by multiplying a factor of 1.8.
We find that the model very well reproduces the observed tight correlation
between the SFR and the stellar mass.

\subsection{Stellar-to-halo mass ratio}
\begin{figure}
    \includegraphics[width=\columnwidth]{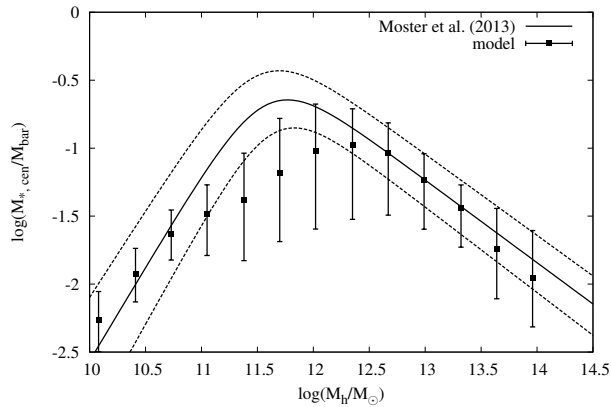}
    \caption{
    The stellar mass of central galaxies relative to the total baryon mass
    in their host halo as a function of the total mass of host halo.
    The black filled squares with error bars denote the median and the 10th 
    to 90th percentile of model galaxies in each bin of the host halo mass.
    The black solid and dashed lines show the average and $1\sigma$ 
    confidence level of stellar mass ratio obtained 
    by abundance matching technique (\citealt{Moster2013}).
    The total baryonic mass $M_{\rm bar}$ is estimated as $M_{\rm bar} =
    M_{\rm h}\times(\Omega_{\rm b}/\Omega_{\rm m})$.
    }
    \label{fig:Mratio}
\end{figure}

Figure \ref{fig:Mratio} presents the ratio of the stellar mass of 
central galaxy to the total baryon mass in their host halo
against the total mass of their host halo.
The total baryon mass $M_{\rm bar}$ is simply estimated as 
$M_{\rm bar} = M_{\rm h}\times(\Omega_{\rm b}/\Omega_{\rm m})$.
This plot indicates an efficiency of star formation as a function of halo mass,
and can be a tight constraint on the galaxy formation model.

The median and the 10th to 90th percentiles of model galaxies in each 
halo mass bin are shown by the black squares with error bars.
The solid and dashed lines show the average and $1 \sigma$ confidence level
estimated by \cite{Moster2013} using an ``abundance matching technique'',
in which the halo mass is estimated by matching the 
abundance of halos in {\it N}-body simulations to the abundance
of observed galaxies.
The prediction of our model well agree with the result of
\cite{Moster2013}.
The distribution of stellar-to-halo mass ratio has a peak
around $M_{\rm h}\sim10^{12}M_{\odot}$.
It reflects effects of SN feedback and AGN feedback:
the former efficiently works in lower mass halos because 
the gravitational potential well is shallow in such halos, 
while the later efficiently works in massive halos because
the cooling time is long enough and the central SMBH can 
sufficiently evolve in such halos.

\subsection{Mass metallicity relation}
\begin{figure}
    \includegraphics[width=\columnwidth]{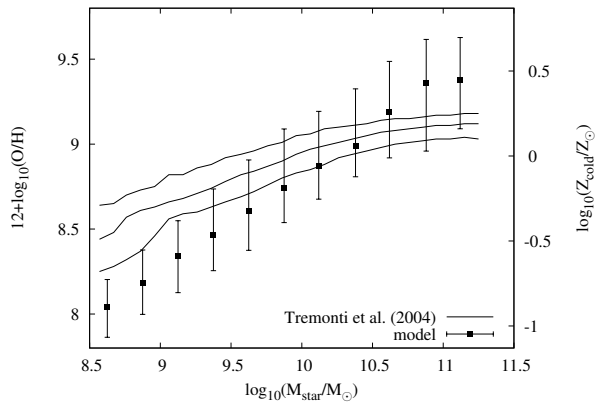}
    \caption{
Relation between stellar mass and cold gas metallicity,
which is denoted by the gas-phase oxygen abundance in 
unit of $12+\log({\rm O/H})$.
Solar metallicity in this unit is 8.93 (Anders \& Grevesse 1989).
The solid lines represents the 84th, 50th, and 16th percentile of local 
star-forming galaxies observed by SDSS (Tremonti et al. 2004).
The black filled squares with error bars show the 84th, 50th, 
and 16th percentile of the distributions of model galaxies in
each magnitude bin.
For the model, we defined star forming galaxy as a galaxy which 
are larger than $10^{-11}$ yr$^{-1}$ in 
specific star formation rate (i.e., SFR/$M_{\rm star}$) .
            }
    \label{fig:MZ}
\end{figure}

Figure \ref{fig:MZ} shows the predicted relation between 
the stellar mass and the metallicity of cold gas for star forming galaxies.
The median and the 16th to 84th percentile of the distribution of 
SDSS galaxies estimated by \cite{Tremonti2004} is also shown in 
figure~\ref{fig:MZ} by solid lines.
The cold gas metallicity is denoted by the gas-phase oxygen
abundance in unit of $12+\log({\rm O/H})$.
The solar metallicity in this unit is $8.93$ \citep{Anders1989}.
The metallicity with respect to the solar metallicity, 
$Z_{\odot}=0.019$ (\citealt{Anders1989}), is also indicated on the
right-side axis of figure~\ref{fig:MZ} for reference.
We defined the ``star-forming galaxy'' as a galaxy with specific
SFR (i.e., SFR/$M_{\rm star}$) higher than $10^{-11} {\rm yr}^{-1}$.
If we change this threshold to more lower value, for example,
the relation will shift towards high-metallicity. 

Comparing with the observation, our model galaxies tend to
have lower metallicities at the stellar mass range of 
$M_{\rm star} < 10^{10} \Msun$.
We will investigate this issue in future paper.

\section{Distant galaxies}
In this section we show the model predictions for the basic properties 
of high-$z$ galaxies.
\label{sec:highzresults}

\subsection{Cosmic star formation history}
\begin{figure}
    \begin{center}
    \includegraphics[width=\columnwidth]{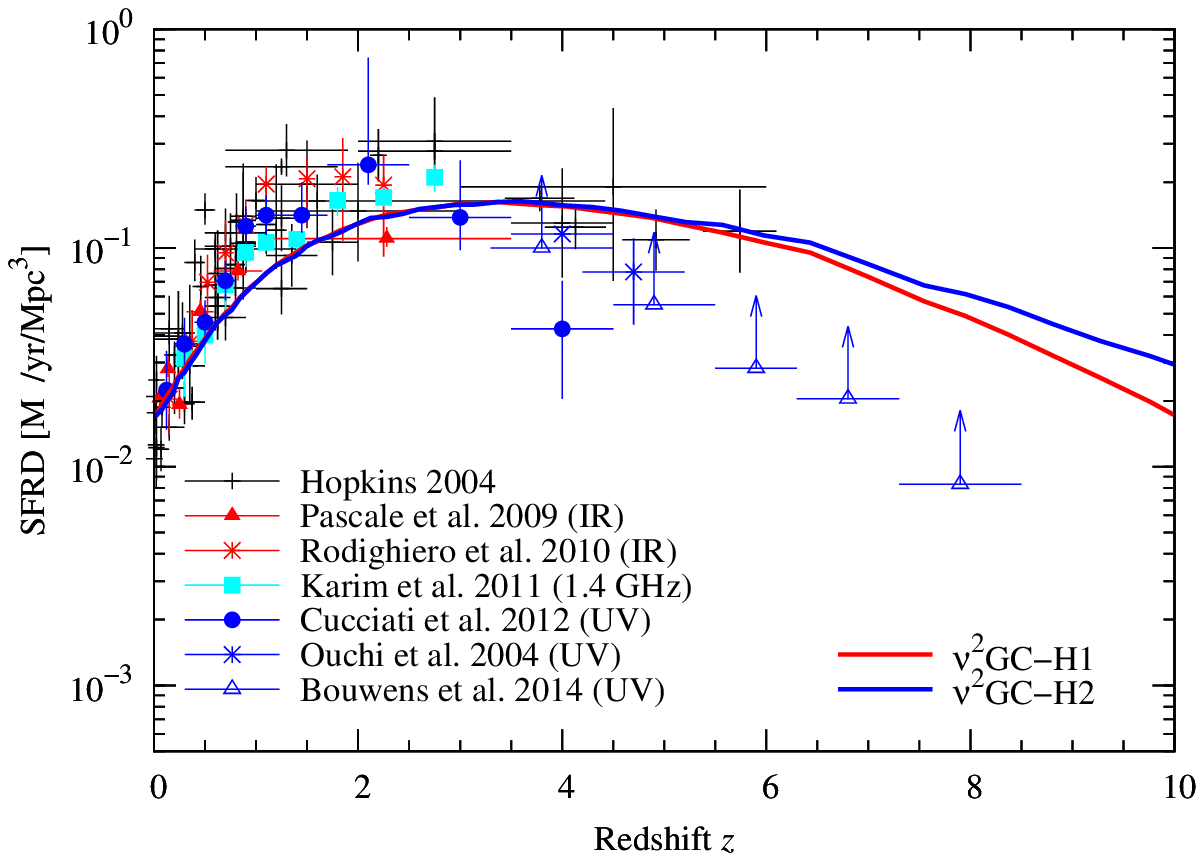}
    \end{center}
    \caption{The cosmic SFR density as a function of redshift.
    The red and blue solid lines represent the predictions by model with
    the {\it N}-body data of $\nu^2$GC-H1 (red) and $\nu^2$GC-H2 (blue), respectively.
    The parameters related to baryon physics are the same in these models.
    We also show the observational data estimated by dust continuum 
    \citep{Pascale2009,Rodighiero2010,Karim2011} and 
    UV continuum \citep{Ouchi2004, Cucciati2012, Bouwens2014}. 
    The data of \cite{Hopkins2004} are the compilation of 
    various observations. All the data points are corrected for dust extinction, 
    by the methods adopted in individual references. 
    The data points of Bouwens et al. (2014) are obtained by integrating 
    LF down to the $M_{\rm AB}$(1500 \AA) $<$ -17.0, while the other observations
    and our model includes the contribution from all galaxies.
    The SFR of model galaxies are converted into those with Salpeter IMF
    from those Chabrier IMF, by multiplying a factor of 1.8.
    }
    \label{fig:csfh}
\end{figure}

Figure \ref{fig:csfh} shows the redshift evolution of cosmic star 
formation rate density (i.e., total SFR of all galaxies per 
unit comoving volume). 
The blue solid line shows the result of our standard model ($\nu^2$GC-H2 model).
The SFR of model galaxies are converted into those with Salpeter IMF
from those Chabrier IMF, by multiplying a factor of 1.8.
The $\nu^2$GC-H1 model is also shown by red solid line to see the effect of
mass resolution.
A discrepancy between these two models increases at high 
redshift, indicating that contributions from galaxies resides 
in lower mass halos become significant at high redshift.

The standard model well reproduces the observations.
At a redshift greater than $z > 4$, it seems that 
the model predictions are much greater than observed SFR density
of \cite{Bouwens2014};
however, their data only include galaxies brighter 
than $M(1500{\rm \AA}) < -17.0$, 
while the other data and model predictions are integration in entire
magnitude range.
Furthermore, the survey of \cite{Bouwens2014} is designed to find 
galaxies having blue colors, and therefore they might miss a population of
dusty red galaxies.
In fact, the predicted UV luminosity density (i.e., total luminosity 
of all galaxies per unit comoving volume) is roughly consistent with
the data of \cite{Bouwens2014} when the effect of limiting magnitude 
are taken into account (see next subsection).

Our model predicts that a large amount of star formation
activity has not yet been observed at distant universe.
It will be investigated by future observations.

\subsection{Evolution of luminosity density in cosmic time}
\begin{figure}
    \includegraphics[width=\columnwidth]{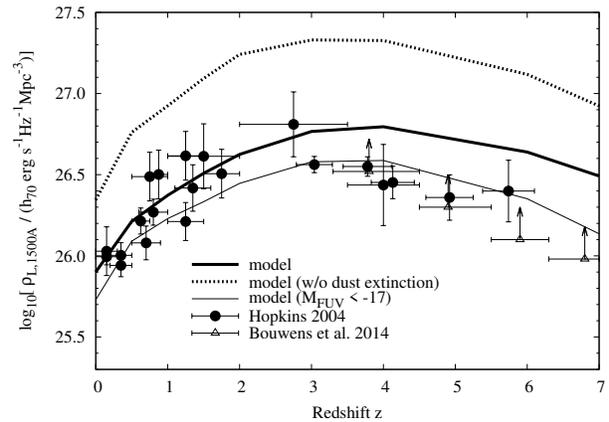}
    \caption{The redshift evolution of luminosity density at 1500 \AA.
    The filled circles and open triangles are observational data compiled by
    \cite{Hopkins2004} and obtained by \cite{Bouwens2014}, respectively.
    The model prediction is shown by solid black line. For the purpose of
    comparison, we also show the model without dust extinction (dotted line).
    Those model predictions include a contribution from all galaxies.
    The data points of \cite{Bouwens2014} are obtained by integrating LF down 
    to the $M_{\rm AB}(1500 \AA) < -17.0$ while the other observational data
    are integrated down to zero luminosity.
    The thin solid line show the model prediction taking into account the 
    magnitude limit of $M_{\rm AB}(1500{\rm \AA}) = -17.0$.}
    \label{fig:uvld}
\end{figure}

Figure \ref{fig:uvld} shows the predicted redshift evolution of 
the luminosity density at 1500 \AA (thick solid line). 
The intrinsic luminosity density (i.e., without dust extinction) 
is shown by thick dotted line  for reference. 
Note that the observational data plotted in figure \ref{fig:uvld} 
are not corrected for dust extinction effect thus they should be 
compared with the model with dust extinction (thick solid line).
As already mentioned above, the data of \cite{Bouwens2014} only include 
 galaxies brighter than $M(1500{\rm \AA}) < -17.0$.
The model prediction taking into account the same magnitude limit with the
\cite{Bouwens2014} is shown by thin solid line.
We can see that the model well reproduces the observations.
This result support a validity of our modelings of star formation
and dust extinction.

\begin{figure}
    \includegraphics[width=\columnwidth]{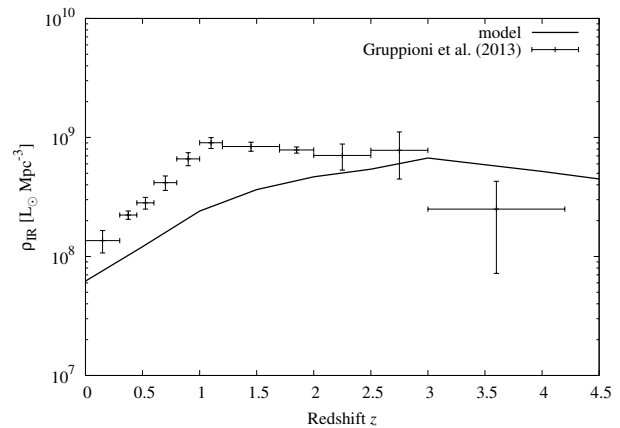}
    \caption{The model prediction for the redshift evolution of total IR luminosity
    density, comparing with the observational data \cite{Gruppioni2013}.
    The total IR luminosity of model galaxies are calculated from the SED of each
    galaxy, to be consistent with the total amount of stellar luminosity
    absorbed by dust.}
    \label{fig:irld}
\end{figure}

Figure \ref{fig:irld} presents the redshift evolution of the 
sum of total IR luminosity ($8$--$1000$ ${\rm \mu m}$) of all galaxies 
per unit comoving volume. 
The total IR luminosity of model galaxies are estimated 
from the SED of each galaxy to be consistent 
with the total amount of stellar luminosity absorbed by dust.
The observational data are obtained by \cite{Gruppioni2013}, 
by integrating the total IR LFs down to $10^8 L_{\odot}$.
The model reproduces the observation within a factor of 2--3.
The discrepancy between the model and observation 
is partly due to a contribution from AGNs, which is included in the 
observational data while not included in the model.

\subsection{Redshift evolution of $K$-band luminosity function}
\begin{figure}
    \begin{center}
    \includegraphics[width=\columnwidth]{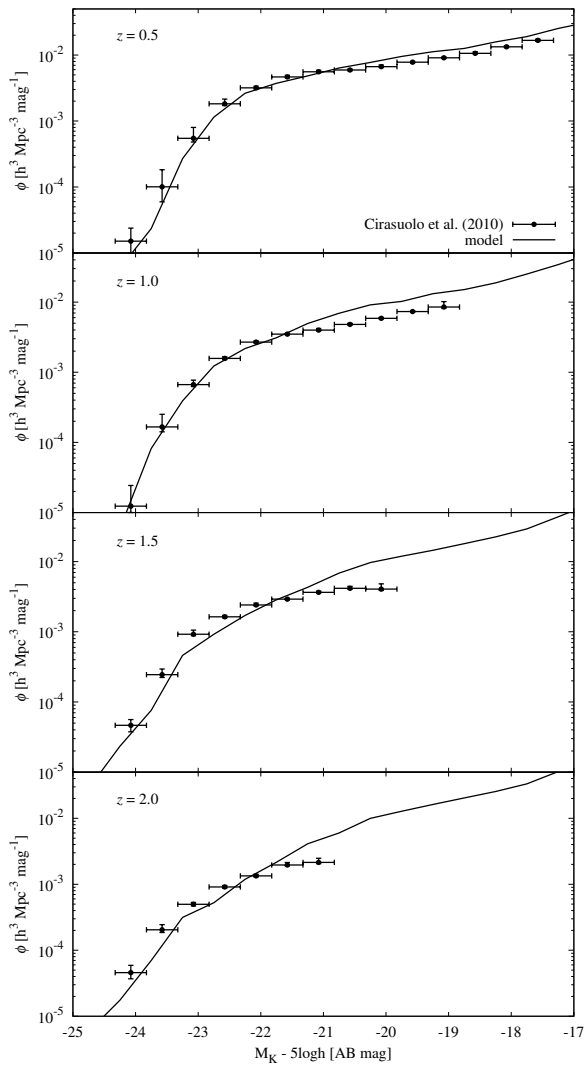}
    \end{center}
    \caption{
    The redshift evolution of rest-frame $K$-band luminosity function.
    From top to bottom, we show the LFs at $z = 0.5, 1.0, 1.5, 2.0$.
    The solid black lines are model predictions.
    The black filled circles with error bars are the observational data obtained 
    by \cite{Cirasuolo2010}.}
    \label{fig:K-evo}
\end{figure}

Figure \ref{fig:K-evo} shows the redshift evolution of rest-frame 
$K$-band LF.
The observational data are obtained by \cite{Cirasuolo2010}.
The model well reproduces the bright-end of LFs even at $z = 2.0$,
which was not able to reproduce in our previous model.
In new model, formation of massive galaxies are suppressed 
by AGN feedback only at low-redshift, and therefore the model
can reproduce the bright-end LFs of local and high-$z$ galaxies at the same time.
On the other hand, the model overestimates the abundance of dwarf galaxies
over all redshift range.
This discrepancy might suggest that SN feedback should be more efficient at high-$z$.
However, there still remains some uncertainties in the observation.
For example, cosmic variance, systematic error in $k$-correction, 
and incompleteness of the survey due to a surface brightness limit
will affect the measurement of the faint-end slope of high-$z$ $K$-band LFs.

\section{Summary}
\label{sec:summary}
In this paper we present a new cosmological galaxy formation model, 
$\nu^2$GC, as an updated version of our previous model, 
$\nu$GC (\citealt{Nagashima2005}; see also \citealt{Nagashima2004}).
Major updates of the model are as follows: 
(1) the {\it N}-body simulations of 
the evolution of dark matter halos are updated (\citealt{Ishiyama2015}), 
(2) The formation and evolution process of SMBHs and 
the suppression of gas cooling due to the AGN activity (AGN feedback)
is included,
(3) heating of the intergalactic gas by the cosmic UV background is included,
and (4) adopt a Markov chain Monte Carlo method in parameter tuning.
Thanks to the updated {\it N}-body simulations, the minimum halo mass
of the model reaches $1.37\times10^8 \Msun$ in the best case, 
which is below the effective Jeans mass at high redshift.
In our largest simulation box (1.12 Gpc/h), 
we can perform statistical analysis for rare objects such as 
bright quasars.

The main results of this paper are summarized as follows.
\begin{enumerate}
\item We tuned the model to fit the local $r$- and $K$-band LFs and 
\HI MF by using a MCMC method. As a result, the model has succeeded 
well in reproducing these observables at the same time.
\item The model well reproduces
the scaling relations between the size and the magnitude, 
and the rotation velocity and the magnitude of spiral galaxies.
For elliptical galaxies, the model roughly well reproduces
the observed size-magnitude relation and the velocity 
dispersion-magnitude relation.
However, for bright elliptical galaxies,
the model underpredicts both of the size and the velocity dispersion.
We need to improve the model related to the galaxy merger and 
formation process of the bulge component.
\item The model well reproduces the observed bimodal distribution 
in color for bright galaxies. On the other hand, the model predicts
redder color for dwarf galaxies comparing with the observations.
This might be caused from our oversimplified prescription
for star formation, SN feedback and stripping of hot gas.
\item For massive galaxies ($M_{\rm star} > 10^{10} \Msun$),
model well reproduces the observed scaling relation between the 
stellar mass and gas phase metallicity at $z = 0$.
However the model underpredicts metallicity of dwarf galaxies.
This might also be caused by our oversimplified treatment of star formation
and SN feedback. In addition, assumed IMF would also affect it.
\item The observed scaling relation between the bulge mass and SMBH mass,
and MF of local SMBHs are well reproduced in our model.
\item The cosmological evolution of star formation rate density
and UV luminosity density predicted by our model are well
agree with the observations.
We found that the model roughly reproduces the redshift evolution of 
total IR luminosity density.
We also compared the redshift evolution of the rest-frame $K$-band LFs,
and found that the model well reproduces the bright-end of LFs at $0 < z < 2$.
\end{enumerate}

Since the main aim of this paper is to present the details of 
the calculation method of our model, 
we compared the model only with some basic observables mentioned above.
Subsequent papers will discuss another topics related to galaxy formation:
the clustering properties of quasars, the origin of cosmic NIR background, 
the properties of sub-millimeter galaxies, for example.

The results of our model, including the LFs in several wavebands, 
mass functions, and the mock galaxies, are publicly available on the web\footnote[1]{}.

\begin{ack}
We would like to thank for the anonymous referee for
many useful comments.
This study has been funded by Yamada Science Foundation, MEXT HPCI
STRATEGIC PROGRAM.
The $\nu^2$GC simulations were partially carried out on the Aterui
supercomputer at theCenter for Computational Astrophysics (CfCA)
of the National Astronomical Observatory of Japan, and the K computer at the RIKEN Advanced Institute for Computational Science
(Proposal numbers hp120286, hp130026, and hp140212).
RM has been supported by the Grant-in-Aid for JSPS Fellows. 
MN has been supported by the Grant-in-Aid for the Scientific Research Fund (25287041) commissioned by the Ministry of Education, Culture, Sports, Science and Technology (MEXT) of Japan.
TO acknowledges the financial support of Japan Society for the Promotion of 
Science (JSPS) Grant-in-Aid for Young Scientists (B: 224740112).
TI has been supported by MEXT/JSPS KAKENHI Grant Number 15K12031.
\end{ack}


\begin{thebibliography}{}

\bibitem[Anders \& Grevesse(1989)]{Anders1989} Anders E., Grevesse N., 
1989, GeCoA,  53, 197 

\bibitem[Avila et al.(2014)]{Avila2014} Avila S., Knebe A., Pearce F.~R., 
et al., 2014, MNRAS,  441, 3488 

\bibitem[Baldry et al.(2004)]{Baldry2004} Baldry I.~K., Glazebrook K., 
Brinkmann J., Ivezi{\'c} {\v Z}., Lupton R.~H., Nichol R.~C., Szalay A.~S., 
2004, ApJ,  600, 681 

\bibitem[Baldry et al.(2012)]{Baldry2012} Baldry, I.~K., Driver, 
S.~P., Loveday, J., et al.\ 2012, \mnras, 421, 621 

\bibitem[Baugh et al.(1996)]{Baugh1996} Baugh C.~M., Cole S., 
Frenk C.~S., 1996, MNRAS,  283, 1361 

\bibitem[Barnes \& Hernquist(1996)]{Barnes96} Barnes, J. E., \& Hernquist, L. 1996, \apj, 471, 115

\bibitem[Bender et al.(1992)]{Bender1992} Bender R., Burstein 
D., Faber S.~M., 1992, ApJ,  399, 462 

\bibitem[Bender \& Nieto(1990)]{Bender1990} Bender R., Nieto J.-L., 1990, 
A\&A,  239, 97

\bibitem[Benson(2014)]{Benson2014} Benson A.~J., 2014, MNRAS,  444, 2599 

\bibitem[Bett et al.(2007)]{Bett2007} Bett P., Eke V., Frenk C.~S., Jenkins 
A., Helly J., Navarro J., 2007, MNRAS,  376, 215

\bibitem[Binney \& Tremaine(1987)]{Binney1987}Binney, J., \& Tremaine,
                                 S. 1987, Galactic Dynamics, Princeton
                                 Univ. Press, Princeton, NJ

\bibitem[Blanton et al.(2005)]{Blanton2005} Blanton M.~R., Lupton R.~H., 
Schlegel D.~J., Strauss M.~A., Brinkmann J., Fukugita M., Loveday J., 2005, 
ApJ,  631, 208 

\bibitem[Boylan-Kolchin et al.(2009)]{Boylan-Kolchin2009} Boylan-Kolchin 
M., Springel V., White S.~D.~M., Jenkins A., Lemson G., 2009, MNRAS,  398, 
1150
    
\bibitem[Bower et al.(2006)]{Bower2006} Bower R.~G., 
Benson A.~J., Malbon R., Helly J.~C., Frenk C.~S., Baugh C.~M., Cole S., 
Lacey C.~G., 2006, MNRAS,  370, 645 
                  
\bibitem[Bouwens et al.(2014)]{Bouwens2014} Bouwens R.~J., Illingworth 
G.~D., Oesch P.~A., et al., 2014, ApJ,  793, 115 
           
\bibitem[Brinchmann et al.(2004)]{Brinchmann2004} Brinchmann J., Charlot 
S., White S.~D.~M., Tremonti C., Kauffmann G., Heckman T., Brinkmann J., 
2004, MNRAS,  351, 1151 
                      
\bibitem[Bullock et al.(2001a)]{Bullock2001}Bullock, J.S., Kolatt, T.S.,
				 Sigad, T., Somerville, R.S., Kravtsov,
				 A.V., Klypin, A.A., Primack, J.R., \&
				 Dekel, A. 2001, \mnras, 321, 559

\bibitem[Bullock et al.(2001b)]{Bullock2001b} Bullock J.~S., Dekel A., Kolatt 
T.~S., Kravtsov A.~V., Klypin A.~A., Porciani C., Primack J.~R., 2001, ApJ,  
555, 240 

\bibitem[Burstein et al.(1997)]{Burstein1997} Burstein D., Bender R., Faber 
S., Nolthenius R., 1997, AJ,  114, 1365 

\bibitem[Bruzual \& Charlot(2003)]{Bruzual2003} Bruzual G., Charlot S., 
2003, MNRAS,  344, 1000 

\bibitem[Calzetti et al.(2000)]{Calzetti2000} Calzetti 
D., Armus L., Bohlin R.~C., Kinney A.~L., Koornneef J., Storchi-Bergmann 
T., 2000, ApJ,  533, 682 

\bibitem[Chabrier(2003)]{Chabrier2003} Chabrier G., 2003, PASP,  115, 763 

\bibitem[Cirasuolo et al.(2010)]{Cirasuolo2010} Cirasuolo, M., 
McLure, R.~J., Dunlop, J.~S., et al.\ 2010, \mnras, 401, 1166
, 11

\bibitem[Cole et al.(1994)]{Cole1994} Cole S., Aragon-Salamanca A., Frenk C.~S., Navarro J.~F., Zepf S.~E., 1994, MNRAS,  271, 781 

\bibitem[Cole et al.(2000)]{Cole2000} Cole S., Lacey C.~G., Baugh C.~M., 
Frenk C.~S., 2000, MNRAS,  319, 168 

\bibitem[Colpi(2014)]{Colpi2014} Colpi M., 2014, SSRv,  183, 189 

\bibitem[Couchman \& Rees(1986)]{Couchman1986} Couchman,
        H.~M.~P., \& Rees, M.~J.\ 1986, \mnras, 221, 53

\bibitem[Courteau et al.(2000)]{Courteau2000} Courteau S., Willick J.~A., 
Strauss M.~A., Schlegel D., Postman M., 2000, ApJ,  544, 636 

\bibitem[Courteau et al.(2007)]{Courteau2007} Courteau S., Dutton A.~A., 
van den Bosch F.~C., MacArthur L.~A., Dekel A., McIntosh D.~H., Dale D.~A., 
2007, ApJ,  671, 203 

\bibitem[Croton et al.(2006)]{Croton2006} Croton, D. J., Springel, 
V., White, S. D. M., et al. 2006, \mnras, 365, 11

\bibitem[Cucciati et al. (2012)]{Cucciati2012} Cucciati, O. et al. 2012, A\&A, 539, A31

\bibitem[Daddi et al.(2007)]{Daddi2007} Daddi E., Dickinson M., Morrison 
G., et al., 2007, ApJ,  670, 156

\bibitem[Dale et al.(1999)]{Dale1999} Dale D.~A., Giovanelli R., Haynes 
M.~P., Campusano L.~E., Hardy E., 1999, AJ,  118, 1489 

\bibitem[Davis et al.(1985)]{Davis1985} Davis M., Efstathiou G., Frenk 
                  C.~S., White S.~D.~M., 1985, \apj, 292, 371 

\bibitem[Davis \& Laor(2011)]{Davis2011} Davis S.~W., Laor A., 2011, ApJ,  
728, 98 

\bibitem[Dekel \& Silk(1986)]{DekelSilk1986} Dekel A., Silk J., 1986, \apj, 303, 39

\bibitem[De Lucia et al.(2010)]{DeLucia2010} De Lucia G., Boylan-Kolchin 
M., Benson A.~J., Fontanot F., Monaco P., 2010, MNRAS,  406, 1533

\bibitem[Di Matteo et al.(2005)]{DiMatteo05}
Di Matteo, T., Springel, V. \& Hernquist, L. 2005, \nat, 7026, 604

\bibitem[Disney et al.(1989)]{ddp89}Disney, M., Davies,
			      J., \& Phillipps, S. 1989, \mnras, 239, 939
			      
\bibitem[Doroshkevich et al.(1967)]{Doroshkevich1967} Doroshkevich,
        A.~G., Zel'dovich, Y.~B., \& Novikov, I.~D.\ 1967,
        \sovast, 11, 233
        
\bibitem[Driver et al.(2012)]{Driver2012} Driver, S.~P., Robotham, 
A.~S.~G., Kelvin, L., et al.\ 2012, \mnras, 427, 3244 

\bibitem[Elahi et al.(2013)]{Elahi2013} Elahi P.~J., Han J., Lux H., et 
al., 2013, MNRAS,  433, 1537 

\bibitem[Elbaz et al.(2007)]{Elbaz2007} Elbaz D., Daddi E., Le Borgne D. et al., 2007, A\&A,  468, 33 

\bibitem[Enoki et al.(2003)]{Enoki2003} Enoki M., Nagashima 
M., Gouda N., 2003, PASJ,  55, 133 

\bibitem[Enoki et al.(2004)]{Enoki2004} Enoki M., Inoue K.~T., Nagashima 
M., Sugiyama N., 2004, ApJ,  615, 19 

\bibitem[Enoki \& Nagashima(2007)]{Enoki2007} Enoki M., Nagashima M., 
2007, PThPh,  117, 241 

\bibitem[Enoki et al.(2014)]{Enoki2014} Enoki M., Ishiyama T., Kobayashi 
M.~A.~R., Nagashima M., 2014, ApJ,  794, 69 

\bibitem[Faber \& Jackson(1976)]{Faber1976} Faber S.~M., Jackson R.~E., 
1976, ApJ,  204, 668

\bibitem[Faber et al.(1989)]{Faber1989} Faber S.~M., Wegner G., Burstein 
D., Davies R.~L., Dressler A., Lynden-Bell D., Terlevich R.~J., 1989, ApJS,  
69, 763 

\bibitem[Fall(1979)]{f79}Fall, S. M. 1979, \nat, 281, 200

\bibitem[Fall \& Efstathiou(1980)]{fe80}Fall, S. M., \& Efstathiou,
                  G. 1980, \mnras, 193, 189

\bibitem[Fall(1983)]{f83}Fall, S. M. 1983, in `Internal kinematics and
                                 dynamics of galaxies', proceedings of
                                 the IAU symposium 100, Besancon,
                                 France, Dordrecht, D. Reidel, p.391
                                 
\bibitem[Firth et al.(2007)]{Firth2007} Firth P., Drinkwater M.~J., 
Evstigneeva E.~A., Gregg M.~D., Karick A.~M., Jones J.~B., Phillipps S., 
2007, MNRAS,  382, 1342 

\bibitem[Forbes et al.(2008)]{Forbes2008} Forbes D.~A., Lasky P., Graham 
A.~W., Spitler L., 2008, MNRAS,  389, 1924 

\bibitem[Gelman \& Rubin (1992)]{Gelman1992} Gelman A., Rubin D., 1992, Stat. Sci., 7, 457

\bibitem[Gnedin(2000)]{Gnedin2000}Gnedin, N. Y. 2000, \apj, 542, 535

\bibitem[Gruppioni et al.(2013)]{Gruppioni2013} Gruppioni C., Pozzi F., 
Rodighiero G., et al., 2013, MNRAS,  432, 23 

\bibitem[Gonzalez-Perez et al.(2014)]{Gonzalez-Perez2014} Gonzalez-Perez 
V., Lacey C.~G., Baugh C.~M., Lagos C.~D.~P., Helly J., Campbell D.~J.~R., 
Mitchell P.~D., 2014, MNRAS,  439, 264 

\bibitem[Hastings(1970)]{Hastings1970} Hastings, W.~K., 1970, Biometrika, 57, 97

\bibitem[Haynes et al.(2011)]{Haynes2011} Haynes M.~P., Giovanelli R., 
Martin A.~M., et al., 2011, AJ,  142, 170 

\bibitem[Henriques et al.(2009)]{Henriques2009} Henriques B.~M.~B., Thomas 
P.~A., Oliver S., Roseboom I., 2009, MNRAS,  396, 535 

\bibitem[Hopkins (2004)]{Hopkins2004} Hopkins 
A.~M., 2004, ApJ,  615, 209 

\bibitem[Hopkins et al.(2005)]{Hopkins05} Hopkins, P.~F., 
Hernquist, L., Cox, T.~J., et al.\ 2005, \apj, 630, 705 

\bibitem[Hopkins et al.(2006)]{Hopkins06} Hopkins, P.~F., 
Hernquist, L., Cox, T.~J., et al.\ 2006, \apjs, 163, 1 

\bibitem[Hopkins et al.(2009)]{Hopkins2009} Hopkins P.~F., Hernquist L., 
Cox T.~J., Keres D., Wuyts S., 2009, ApJ,  691, 1424 

\bibitem[Ishiyama et al.(2009)]{Ishiyama2009} Ishiyama T., 
Fukushige T., Makino J., 2009, PASJ,  61, 1319 
				 
\bibitem[Ishiyama et al.(2012)]{Ishiyama2012}
Ishiyama, T., Nitadori, K., Makino, J. 2012, in Proc. Int. Conf. High
  Performance Computing, Networking, Storage and Analysis, SC'12 (Los Alamitos,
  CA: IEEE Computer Society Press), 5:, (arXiv:1211.4406)

\bibitem[Ishiyama et al.(2015)]{Ishiyama2015} Ishiyama T., Enoki M., 
Kobayashi M.~A.~R., Makiya R., Nagashima M., Oogi T., 2015, PASJ,  67, 61 

\bibitem[Jaffe(1983)]{jaffe}Jaffe, W. 1983, \mnras, 202, 995

\bibitem[Jiang et al.(2008)]{Jiang2008} Jiang C.~Y., Jing Y.~P., 
Faltenbacher A., Lin W.~P., Li C., 2008, ApJ,  675, 1095 

\bibitem[Jiang et al.(2010)]{Jiang2010} Jiang C.~Y., Jing Y.~P., 
Lin W.~P., 2010, A\&A,  510, A60 

\bibitem[Karim et al. (2011)]{Karim2011} Karim, A. et al. 2011, ApJ, 730, 61

\bibitem[Kauffmann \& White(1993)]{Kauffmann1993a} Kauffmann G., White 
S.~D.~M., 1993, MNRAS,  261, 921 

\bibitem[Kauffmann et al.(1993b)]{Kauffmann1993b} Kauffmann G., White S.~D.~M., Guiderdoni B., 1993, MNRAS,  264, 201

\bibitem[Keres et al.(2003)]{Keres2003} Keres D., Yun M.~S., Young 
J.~S., 2003, ApJ,  582, 659 

\bibitem[Knebe et al.(2011)]{Knebe2011} Knebe A., Knollmann S.~R., Muldrew 
S.~I., et al., 2011, MNRAS,  415, 2293 

\bibitem[Knebe et al.(2015)]{Knebe2015} Knebe A., Pearce F.~R., Thomas 
P.~A., et al., 2015, MNRAS,  451, 4029 

\bibitem[Kobayashi et al.(2007)]{Kobayashi2007} Kobayashi 
M.~A.~R., Totani T., Nagashima M., 2007, ApJ,  670, 919 

\bibitem[Kobayashi et al.(2010)]{Kobayashi2010} Kobayashi 
M.~A.~R., Totani T., Nagashima M., 2010, ApJ,  708, 1119 

\bibitem[Koyama et al.(2008)]{Koyama2008} Koyama H., Nagashima M., Kakehata T., Yoshii Y., 2008, \mnras, 389, 237

\bibitem[Kuzmin(1952)]{kuzmin1952} Kuzmin G., 1952, Publ. Astron. Obs. Tartu, 32, 311

\bibitem[Kuzmin(1956)]{kuzmin1956} Kuzmin G., 1956, Astron. Zh., 33, 27

\bibitem[Lacey \& Cole(1993)]{Lacey1993}Lacey, C.G., \& Cole, S. 1993,
                                 \mnras, 262, 627   

\bibitem[Lagos et al.(2014)]{Lagos2014} Lagos C.~d.~P., Davis T.~A., Lacey 
C.~G., Zwaan M.~A., Baugh C.~M., Gonzalez-Perez V., Padilla N.~D., 2014, 
MNRAS,  443, 1002 

\bibitem[Lee et al.(2014)]{Lee2014} Lee J., Yi S.~K., Elahi P.~J., et al., 
2014, MNRAS,  445, 4197 

\bibitem[Lu et al.(2012)]{Lu2012} Lu, Y., Mo, H.~J., Katz, N., 
\& Weinberg, M.~D.\ 2012, \mnras, 421, 1779 

\bibitem[Lu et al.(2014)]{Lu2014} Lu, Y., Mo, H.~J., Lu, Z., 
Katz, N., \& Weinberg, M.~D.\ 2014, \mnras, 443, 1252 

\bibitem[Macci{\`o} et al.(2008)]{Maccio2008} 
Macci{\`o} A.~V., Dutton A.~A., van den Bosch F.~C., 2008, MNRAS,  391, 
1940 

\bibitem[Madau \& Dickinson(2014)]{Madau2014} Madau P., Dickinson M., 
2014, ARA\&A,  52, 415 

\bibitem[Maeder(1992)]{Maeder1992} Maeder, A. 1992, \aap, 264, 105

\bibitem[Makino \& Hut(1997)]{Makino1997}Makino, J., \& Hut, P. 1997, \apj,
                                 481, 83

\bibitem[Makiya et al.(2011)]{Makiya2011} Makiya R., Totani 
T., Kobayashi M.~A.~R., 2011, ApJ,  728, 158 

\bibitem[Makiya et al.(2014)]{Makiya2014} Makiya R., Totani T., Kobayashi 
M.~A.~R., Nagashima M., \& Takeuchi T.~T., 2014, MNRAS,  441, 63 

\bibitem[Martin et al.(2010)]{Martin2010} Martin A.~M., Papastergis E., 
Giovanelli R., Haynes M.~P., Springob C.~M., Stierwalt S., 2010, ApJ,  723, 
1359 

\bibitem[Mathewson et al.(1992)]{Mathewson1992} Mathewson 
D.~S., Ford V.~L., Buchhorn M., 1992, ApJS,  81, 413 

\bibitem[Matkovi\'c \& Guzm\'an(2005)]{Matkovic2005} Matkovi{\'c} 
A., Guzm{\'a}n R., 2005, MNRAS,  362, 289 

\bibitem[McConnell \& Ma (2013)]{McConnell13}
McConnell, N.~J. \& Ma, C.-P., 2013, \apj, 764, 181

\bibitem[Metropolis et al.(1953)]{Metropolis1953} Metropolis, N., 
Rosenbluth, A.~W., Rosenbluth, M.~N., Teller, A.~H., 
\& Teller, E.\ 1953, \jcp, 21, 1087 

\bibitem[Mihos \& Hernquist(1994)]{Mihos1994} Mihos, J. C., \& Hernquist, L. 1994, \apj, 431, L9 

\bibitem[Mihos \& Hernquist(1996)]{Mihos1996} Mihos, J. C., \& Hernquist, L. 1996, \apj, 464, 641

\bibitem[Mo et al.(1998)]{mmw98}Mo, H.J., Mao, S., \& White,
                                 S.D.M. 1998, \mnras, 295, 319

\bibitem[Monaco et al.(2007)]{Monaco2007} Monaco P., 
Fontanot F., Taffoni G., 2007, MNRAS,  375, 1189 

\bibitem[Monaco et al.(2014)]{Monaco2014} Monaco P., Benson A.~J., De Lucia 
G., Fontanot F., Borgani S., Boylan-Kolchin M., 2014, MNRAS,  441, 2058

\bibitem[Moore et al.(2002)]{Moore2002} Moore S.~A.~W., Lucey J.~R., 
Kuntschner H., Colless M., 2002, MNRAS,  336, 382 

\bibitem[Moster et al.(2013)]{Moster2013} Moster B.~P., Naab T., 
White S.~D.~M., 2013, MNRAS,  428, 3121 

\bibitem[Nagashima \& Yoshii(2003)]{Nagashima2003}Nagashima, M., \& Yoshii,
			      Y. 2003, \mnras, 340, 509
			      
\bibitem[Nagashima \& Yoshii(2004)]{Nagashima2004} Nagashima M., Yoshii 
Y., 2004, ApJ,  610, 23

\bibitem[Nagashima et al.(2005)]{Nagashima2005} Nagashima M., Yahagi H., Enoki 
M., Yoshii Y., Gouda N., 2005, ApJ,  634, 26 (N05)

\bibitem[Navarro, Frenk \& White(1997)]{nfw97} Navarro J. F., Frenk C. S., White S. D. M., 1997, \mnras, 490, 493

\bibitem[Okamoto \& Habe(1999)]{Okamoto1999} Okamoto, T., \& Habe, A. 1999, \apj, 516, 591

\bibitem[Okamoto \& Habe(2000)]{Okamoto2000} Okamoto, T., \& Habe, A. 2000, \pasj, 52, 457
      
\bibitem[Okamoto et al.(2008)]{Okamoto2008} Okamoto, T., Gao, L., \& Theuns, T.\ 2008, \mnras, 390, 920

\bibitem[Okoshi \& Nagashima(2005)]{Okoshi2005} Okoshi K., Nagashima M., 
2005, ApJ,  623, 99 

\bibitem[Okoshi et al.(2010)]{Okoshi2010} Okoshi K., Nagashima M., Gouda 
N., Minowa Y., 2010, ApJ,  710, 1295 

\bibitem[Onions et al.(2012)]{Onions2012} Onions J., Knebe A., Pearce 
F.~R., et al., 2012, MNRAS,  423, 1200 

\bibitem[Oogi et al.(2016)]{Oogi2016} Oogi T., Enoki M., Ishiyama T., 
Kobayashi M.~A.~R., Makiya R., Nagashima M., 2016, MNRAS,  456, L30 

\bibitem[Ouchi et al. (2004)]{Ouchi2004} Ouchi, M. et al. 2004, ApJ, 611, 685

\bibitem[Pascale et al. (2009)]{Pascale2009} Pascale, E. et al. 2009, ApJ, 707, 1740

\bibitem[Planck Collaboration et al.(2014)]{Planck2014} 
Planck Collaboration, Ade P.~A.~R., Aghanim N., et al., 2014, 
A\&A,  571, A16 

\bibitem[Power et al.(2010)]{Power2010} Power, C., Baugh, C.~M., 
\& Lacey, C.~G.\ 2010, \mnras, 406, 43

\bibitem[Prada et al.(2012)]{Prada2012} Prada F., Klypin A.~A., Cuesta 
A.~J., Betancort-Rijo J.~E., Primack J., 2012, MNRAS,  423, 3018 

\bibitem[Rodighiero et al. (2010)]{Rodighiero2010} Rodighiero, G. et al. 2010, A\&A, 515, A8

\bibitem[Roukema et al.(1997)]{Roukema1997} Roukema B.~F., Quinn P.~J., Peterson B.~A., Rocca-Volmerange B., 1997, MNRAS,  292, 835 

\bibitem[Salim et al.(2007)]{Salim2007} Salim S., Rich R.~M., Charlot S., et al., 2007, ApJS,  173, 267 

\bibitem[S{\'a}nchez-Conde \& Prada(2014)]{SanchezConde2014} 
S{\'a}nchez-Conde M.~A., Prada F., 2014, MNRAS,  442, 2271 

\bibitem[Shankar et al.(2004)]{Shankar2004} Shankar F., Salucci P., Granato 
G.~L., De Zotti G., Danese L., 2004, MNRAS,  354, 1020 

\bibitem[Shankar et al.(2013)]{Shankar2013} Shankar F., Marulli F., 
Bernardi M., Mei S., Meert A., Vikram V., 2013, MNRAS,  428, 109 

\bibitem[Shimizu et al.(2002)]{Shimizu2002}Shimizu, M., Kitayama, T., Sasaki,
				 S., \& Suto, Y. 2002, \pasj, 54, 645
				
\bibitem[Shirakata et al.(2015)]{Shirakata2015} Shirakata H., Okamoto T., Enoki M., Nagashima M., Kobayashi M.~A.~R., Ishiyama T., Makiya R., 2015, MNRAS,  450, L6 

\bibitem[Simien \& de Vaucouleurs(1986)]{Simien1986} Simien F., de 
Vaucouleurs G., 1986, ApJ,  302, 564 

\bibitem[Springel et al.(2005)]{Springel2005} Springel V., White S.~D.~M., 
Jenkins A., et al., 2005, Natur,  435, 629 
	 
\bibitem[Springel(2012)]{Springel2012} Springel V., 2012, AN,  333, 515

\bibitem[Somerville \& Primack(1999)]{Somerville1999} Somerville R.~S., Primack J.~R., 1999, MNRAS,  310, 1087 

\bibitem[Somerville(2002)]{Somerville2002} Somerville R.~S., 2002, ApJ,  
572, L23 

\bibitem[Somerville \& Dav{\'e}(2014)]{Somerville2014} 
  Somerville R.~S., Dav{\'e} R. \ 2014, arXiv,  arXiv:1412.2712 

\bibitem[Srisawat et al.(2013)]{Srisawat2013} Srisawat C., Knebe A., Pearce 
F.~R., et al., 2013, MNRAS,  436, 150 

\bibitem[Sutherland \& Dopita(1993)]{Sutherland1993}Sutherland, R., \& Dopita,
                                  M. A. 1993, \apjs, 88, 253

\bibitem[Trager et al.(2000)]{Trager2000} Trager S.~C., Faber S.~M., 
Worthey G., Gonz{\'a}lez J.~J., 2000, AJ,  120, 165 

\bibitem[Tremonti et al.(2004)]{Tremonti2004} Tremonti C.~A., Heckman 
T.~M., Kauffmann G., et al., 2004, ApJ,  613, 898 

\bibitem[Tully \& Fisher(1977)]{Tully1977} Tully R.~B., Fisher J.~R., 
1977, A\&A,  54, 661

\bibitem[Tully et al.(1996)]{Tully1996} Tully R.~B., Verheijen M.~A.~W., 
Pierce M.~J., Huang J.-S., Wainscoat R.~J., 1996, AJ,  112, 2471 

\bibitem[Verheijen(2001)]{Verheijen2001} Verheijen M.~A.~W., 2001, ApJ,  
563, 694 

\bibitem[Wetzel(2011)]{Wetzel2011} Wetzel, A.~R.\ 2011, \mnras, 412, 49 

\bibitem[White \& Frenk(1991)]{White1991} White S.~D.~M., Frenk C.~S., 
1991, ApJ,  379, 52 

\bibitem[Yoshii \& Arimoto(1987)]{ya87} Yoshii Y., Arimoto N., 1987, A\&A, 188, 13

\bibitem[Zwaan et al.(2005)]{Zwaan2005} Zwaan M.~A., Meyer M.~J., 
Staveley-Smith L., Webster R.~L., 2005, MNRAS,  359, L30 




\end{thebibliography}
\end{document}